\documentclass[epj,nopacs,fleqn]{svjour}
\pdfoutput=1
\usepackage{slashed,graphicx,amsmath,amssymb,cite,url,hyperref}
\usepackage[utf8]{inputenc}
\hypersetup{bookmarks=true,unicode=true,pdftoolbar=true,pdfmenubar=true,
  pdffitwindow=false,pdfstartview={FitH},
  pdfnewwindow=true,colorlinks=true,
  linkcolor=blue,citecolor=magenta,filecolor=magenta,urlcolor=cyan}

\newcommand\sss{\scriptscriptstyle}
\newcommand{\mW}{m_{\sss W}}

\def\bsp#1\esp{\begin{split}#1\end{split}}
\def\bpm{\begin{pmatrix}}
\def\epm{\end{pmatrix}}

\newcommand{\bea}{\begin{eqnarray}}  
\newcommand{\eea}{\end{eqnarray}}  
\newcommand{\be}{\begin{equation}}
\newcommand{\ee}{\end{equation}}
\def\cg{\tilde c_g}
\def\cga{\tilde c_\gamma}
\def\ctw{\tilde c_{3W}}
\def\chw{\tilde c_{HW}}
\def\chb{\tilde c_{HB}}

\begin{document}

\title{Probing $CP$-violating Higgs and gauge boson couplings in the Standard
Model effective field theory}

\date{}

\author{
  Felipe Ferreira\inst{1,2,}\thanks{\color{blue}felipefreitas@fisica.ufpb.br},
  Benjamin Fuks\inst{3,4,5,}\thanks{\color{blue}fuks@lpthe.jussieu.fr},
  Ver\'onica Sanz\inst{1,}\thanks{\color{blue}v.sanz@sussex.ac.uk},
  Dipan Sengupta\inst{6,7,}\thanks{\color{blue}dipan@lpsc.in2p3.fr}
}

\institute{
  Department of Physics and Astronomy, University of Sussex,
  Brighton BN1 9QH, UK
  \and
  Departamento de F\'isica, Universidade Federal da Para\'iba,
  Caixa Postal 5008, 58051-970, Jo\~ao Pessoa, Para\'iba, Brazil
  \and
  Sorbonne Universit\'es, Universit\'e Pierre et Marie Curie
  (Paris 06), UMR 7589, LPTHE, F-75005, Paris, France
  \and
  CNRS, UMR 7589, LPTHE, F-75005, Paris, France
  \and
  Institut Universitaire de France, 103 boulevard Saint-Michel,
  75005 Paris, France
  \and
  Laboratoire de Physique Subatomique et de Cosmologie, Universit\'e
  Grenoble-Alpes, CNRS/IN2P3, Avenue des Martyrs 53, F-38026 Grenoble,
  France
  \and
  Department of Physics and Astronomy, Michigan State University,
  East Lansing, U.S.A
}

\abstract{We study the phenomenological consequences of several $CP$-violating
  structures that could arise in the Standard Model effective field theory
  framework. Focusing on operators involving electroweak gauge and/or Higgs
  bosons, we derive constraints originating from Run~I LHC data. We then study
  the capabilities of the present and future LHC runs at higher energies to
  further probe associated $CP$-violating phenomena and we demonstrate how
  differential information can play a key role. We consider both
  traditional four-lepton probes of $CP$-violation in the Higgs sector and novel
  new physics handles based on varied angular and non-angular observables.}

\titlerunning{Probing CPV Higgs and gauge boson couplings in the SMEFT}
\authorrunning{F.~Ferreira {\it et al.}}

\maketitle
\flushbottom

\section{Introduction}
While the discovery of the 125~GeV Higgs boson~\cite{Aad:2012tfa,%
Chatrchyan:2012xdj} has been an emphatic triumph of the first run of the LHC,
questions about the true nature of the new boson still persist. The measured
properties of the Higgs boson are so far consistent with the Standard Model
predictions within the margins of the theoretical and experimental
uncertainties~\cite{Khachatryan:2016vau}, but current data still leaves enough
room for deviations. As a consequence, one of the main topics of
the next LHC runs consists of precisely measuring the Higgs boson properties,
\textit{i.e.}, its couplings to the Standard Model particles and its $CP$
nature.

One of the simplest model-independent way of analyzing deviations from the
Standard Model in the properties of the Higgs boson relies on the Effective
Field
Theory (EFT) language. In this approach, all new physics contributions to the
Standard Model are parameterized in terms of higher-dimensional operators, the
corresponding Wilson coefficients encoding the dependence on the ultraviolet
completion of the Standard Model being taken as free parameters. The EFT
approach can
be tested {\it per se} by investigating the correlations among the
signatures expected both at the LHC and in low-energy experiments, which
equivalently constrains the allowed range for the Wilson coefficients in the
light of current data. Focusing on the possibly $CP$-violating nature of the
Higgs boson interactions, data is currently consistent with a $CP$-even
hypothesis, like in the Standard Model. There however still exists a large
fraction of the Wilson coefficient parameter space where the Higgs boson could
exhibit $CP$-odd couplings to vector bosons and fermions. While this regions is
mostly phenomenologically and experimentally unexplored, it remains important
for model building considerations, as new sources of $CP$ violation (CPV)
are necessary to realize electroweak
baryogenesis~\cite{Espinosa:2011eu}.

The impact of higher-dimensional operators modifying the way in which
the Higgs boson interacts with the electroweak bosons has been extensively
probed in the past. Most studies however assume that the new physics
contributions to the Higgs boson couplings feature a $CP$-even structure,
in particular when existing constraints on the
effective operators are evaluated~\cite{Corbett:2012ja,Dumont:2013wma,%
deBlas:2014ula,Falkowski:2014tna,Dumont:2014lca,Ellis:2014jta,Ellis:2014dva,%
Butter:2016cvz}. In comparison, the investigation of the effects of the $CP$-odd
Higgs boson effective operators has been relatively sparse~\cite{Dawson:2013bba,Anderson:2013afp,Delaunay:2013npa,
Dwivedi:2015nta,Gritsan:2016hjl,Khatibi:2014bsa,Dwivedi:2016xwm}, although some experimental analyses are available, e.g.~\cite{Khachatryan:2014kca,Khachatryan:2016tnr}. As far as gauge interactions are concerned, CPV effects can be
parameterized by six independent dimensions-six operators yielding novel
interactions involving at least either three gauge and Higgs bosons, or gauge
bosons only. The magnitude of the corresponding Wilson coefficients is in
general constrained by electric dipole moments data and electroweak precision
tests~\cite{Weinberg:1989dx,Eidelman:2004rpw,Dwivedi:2015nta,Chien:2015xha}, as well as by
 fits of Higgs coupling measurements at the LHC~\cite{Chatrchyan:2012jja,%
Manohar:2006gz,Chang:2013cia,Belusca-Maito:2014dpa,Belusca-Maito:2015lna}.

In the light of the amount of LHC data to be recorded in the
following years, it is important to consider both options of CPV and $CP$-%
conserving new physics Higgs-boson interactions. The discrimination between
these two kinds of effects is however only achievable once suitable observables
allowing us to probe the $CP$ nature of the Higgs couplings are considered.
Pioneering works have followed this path and investigated
handles that can be obtained from the study of asymmetries in
specific observables~\cite{Choi:2002jk,Godbole:2007cn,Hagiwara:2009wt,%
Englert:2010ud,Dwivedi:2015nta}. Effective scales $\Lambda$ that range up to
40~TeV have been found to be reachable with an LHC integrated luminosity of
about 3000~fb$^{-1}$, assuming ${\cal O}(1)$ Wilson coefficients.

The performed studies are however far from being exhaustive, both in terms of
the
considered set of differential distributions and the Higgs production and decay
channels scrutinized. A significant number of other potential appealing options
have indeed been left over, and could be used to unravel a potential
$CP$-odd nature of the Higgs boson.
In this paper, we focus on a dedicated set of observables that allows us to get
a better handle on the CPV operators by studying several electroweak Higgs boson
production processes, as pointed out in the context of the LHC Higgs Cross
Section Working Group~\cite{deFlorian:2016spz}. We first consider dimensionful
quantities
for which the high-energy regime is automatically sensitive to the large
momentum transfers induced by the EFT operators. We next consider angular
observables that are naturally sensitive to the $CP$-violating nature of the
considered operator. The complete quantitative analysis of this joint effect is
left for future works.

The rest of the paper is organized as follows. In Section~\ref{sec:EFT}, we present the effective
Lagrangian that we have used as a benchmark model, and we briefly discuss its
possible connection to ultraviolet-complete extensions of the Standard Model
in Section~\ref{sec:UV}. In Section~\ref{sec:Run1}, we make use of the LHC Run~I
data to define the region of the Wilson coefficient parameter space that is
relevant for the Run~II studies that we have performed.
Section~\ref{sec:future} is dedicated to prospects arising from the use of
total rates only, and Section~\ref{sec:futurediffs} focuses on differential
kinematic information. Our results are summarized and discussed in
Section~\ref{sec:discussion} and
Section~\ref{sec:conclusions}.

\section{Effective field theory framework}\label{sec:EFT}
In the Standard Model EFT framework, all new physics effects are parameterized
by means of higher-dimensional operators involving the Standard Model fields and
assumed to stem from new phenomena occurring at a large energy scale $\Lambda$.
Considering that the leading effects of physics beyond the Standard Model are
described by operators of dimension six $\{ {\cal O}_i \}$, the Lagrangian
modelling our theoretical framework is given by
\be
  {\cal L}^{(6)}_{\rm EFT} = {\cal L}_{\rm SM}
    + \sum_i \frac{\tilde c_i}{\mW^2} {\cal O}_{i}\ ,
\ee
where ${\cal L}_{\rm SM}$ stands for the Standard Model Lagrangian. In the
above expression, we have normalized the Wilson coefficients $\tilde c$ in a way
in which the effective scale $\Lambda$ is identified with the $W$-boson mass
$\mW$.

\begin{figure}
  \centering
  \includegraphics[width=0.485\columnwidth]{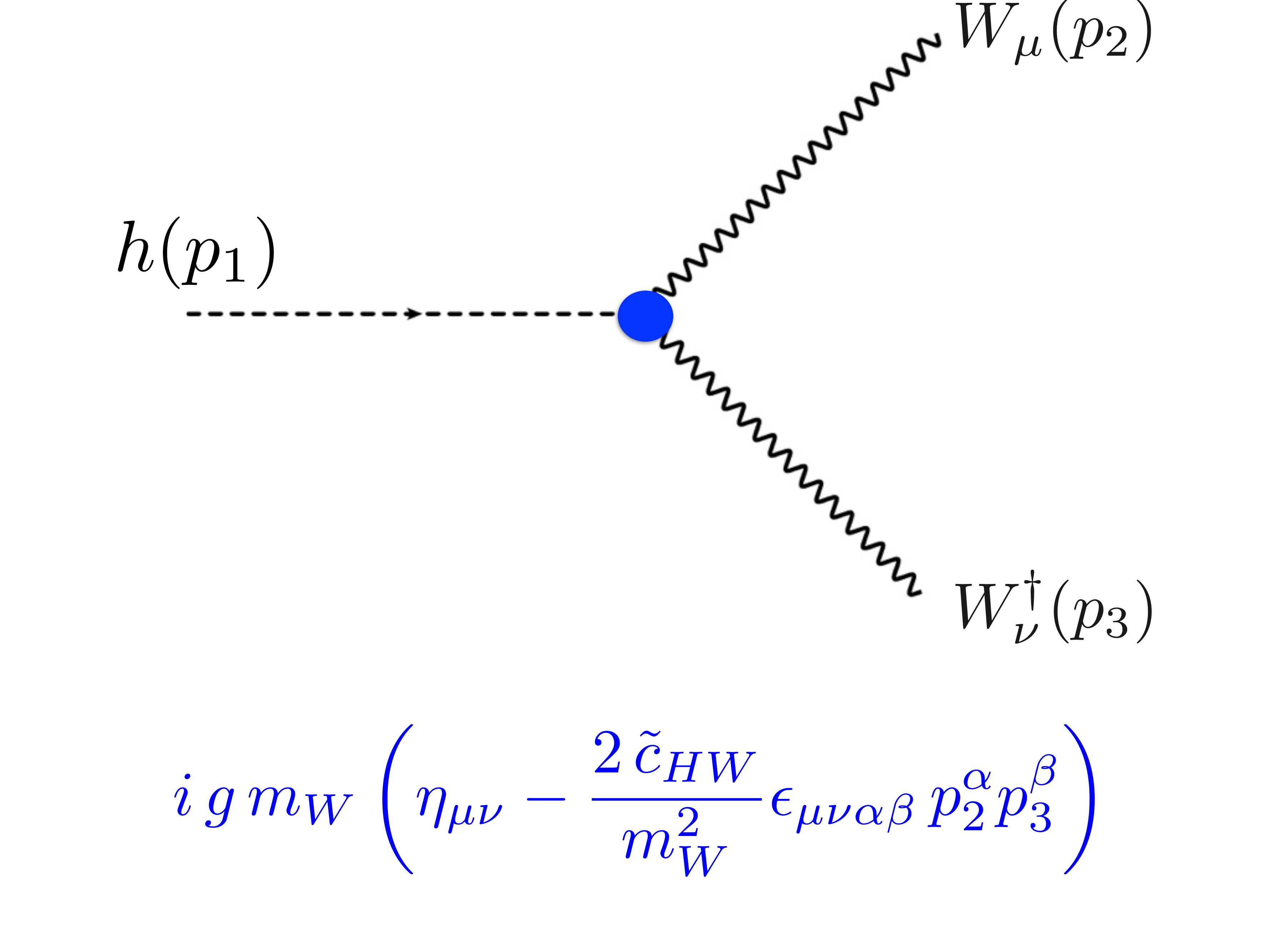}
  \includegraphics[width=0.485\columnwidth]{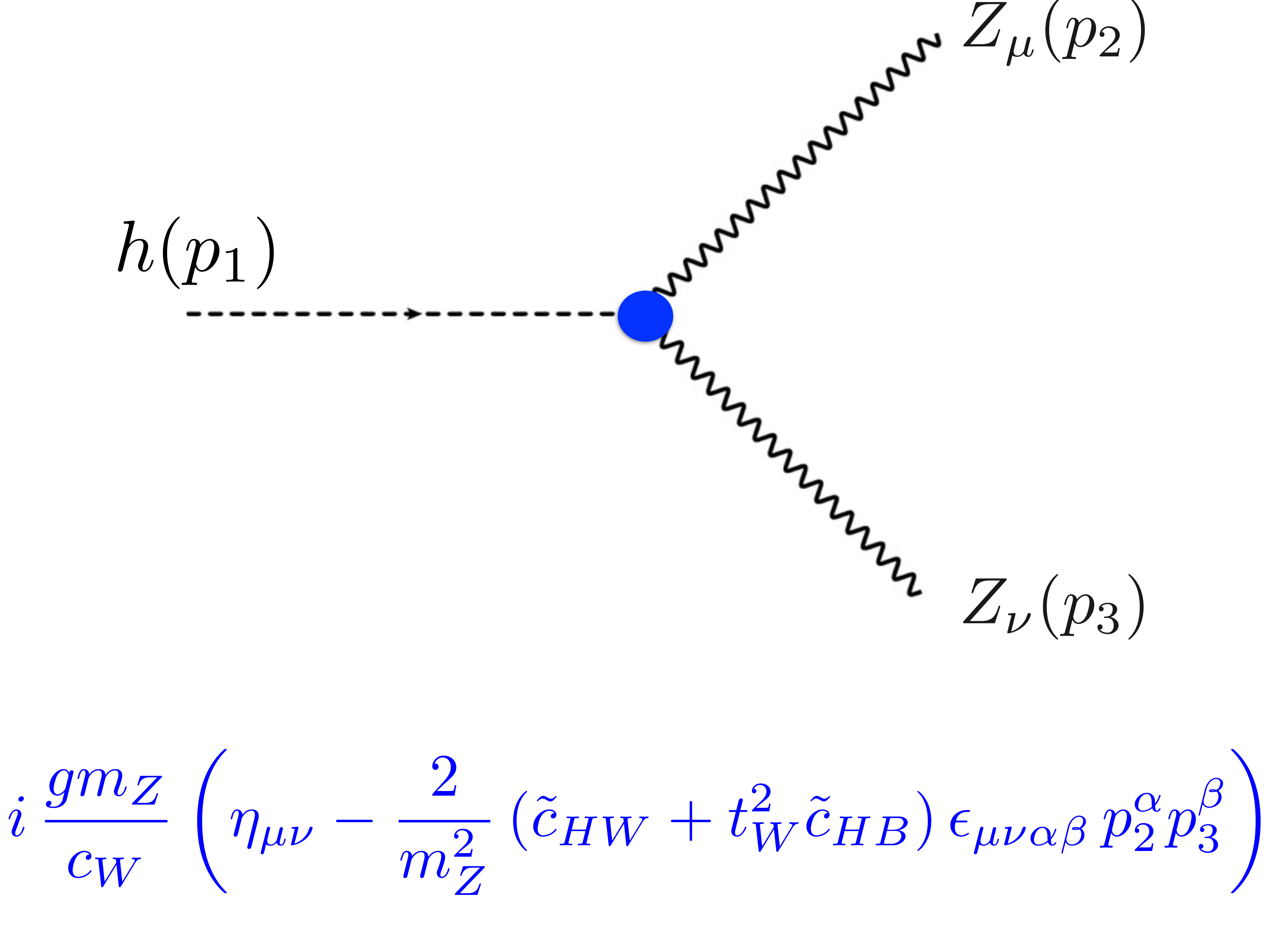}
  \caption{Feynman rules associated with dimension-six CPV operators involving
    a Higgs boson and a pair of weak bosons.}\label{fig:cpvFR}
\end{figure}

The most general ${\cal L}^{(6)}_{\rm EFT}$ Lagrangian invariant under the
Standard Model $SU(3)_c \times SU(2)_L \times U(1)_Y$ gauge symmetries is known
for a long time~\cite{Burges:1983zg,Leung:1984ni,Buchmuller:1985jz}, and is
usually casted in a suitable form by adopting a convenient basis of independent
operators~\cite{Giudice:2007fh,Grzadkowski:2010es,Contino:2013kra,%
Gupta:2014rxa}. In this work, we focus on the dimension-six CPV interactions of
the Higgs and the electroweak gauge bosons that are written, in a form inspired
by the SILH basis conventions~\cite{Giudice:2007fh,Contino:2013kra}, as
\be \label{LCPV}\bsp
& {\cal L}_{\rm CP} =
    i g \frac{\tilde c_{\sss HW}}{\mW^2}  D^\mu \Phi^\dag T_{2k} D^\nu \Phi
     {\widetilde W}_{\mu \nu}^k
  + i g' \frac{\tilde c_{\sss HB}}{\mW^2} D^\mu \Phi^\dag D^\nu \Phi
     {\widetilde B}_{\mu \nu}\\
  &\ \
  + g'^2 \frac{\tilde c_{\sss \gamma}}{\mW^2} \Phi^\dag \Phi B_{\mu\nu}
     {\widetilde B}^{\mu\nu}
  + g_s^2 \frac{\tilde c_{\sss g}}{\mW^2} \Phi^\dag \Phi G_{\mu\nu}^a
     {\widetilde G}^{\mu\nu}_a\\ &\ \
  + g^3 \frac{\tilde c_{\sss 3W}}{\mW^2} \epsilon_{ijk} W_{\mu\nu}^i
      W^\nu{}^j_\rho {\widetilde W}^{\rho\mu k}
  + g_s^3 \frac{\tilde c_{\sss 3G}}{\mW^2} f_{abc} G_{\mu\nu}^a G^\nu{}^b_\rho
     {\widetilde G}^{\rho\mu c} \ ,
\esp \ee
where $B_{\mu\nu}$, $W_{\mu\nu}$ and $G_{\mu\nu}$ ($\widetilde B_{\mu\nu}$,
$\widetilde W_{\mu\nu}$ and $\widetilde G_{\mu\nu}$) denote the hypercharge,
weak isopsin and strong (dual) field strength tensors respectively. In addition,
$\Phi$ represents the electroweak doublet of Higgs fields, $g'$, $g$ and $g_s$
are the $SU(3)_c$, $SU(2)_L$ and $U(1)_Y$ gauge coupling constants and
$\epsilon_{ijk}$ and $f_{abc}$ are the $SU(2)$ and $SU(3)$ group structure
constants. Translations of the ${\cal L}_{\rm CP}$ Lagrangian into any other
commonly considered bases~\cite{Alloul:2013naa,Alonso:2013hga,deFlorian:2016spz}
can be automatically performed with, {\it e.g.}, the {\sc Rosetta}
package~\cite{Falkowski:2015wza}.

The ${\cal L}_{\rm CP}$ Lagrangian induces new Lorentz structures, such as those
featured in the Feynman rules depicted in Figure~\ref{fig:cpvFR}, that have a
manifest CPV structure. Although the restricted set of operators included in
Eq.~\eqref{LCPV} can in principle be extended by CPV fermionic operators~\cite{%
Berge:2008wi,Berge:2008dr,Berge:2011ij,Brod:2013cka,Harnik:2013aja}, we
postpone the study of the latter to a future work. We moreover consider
observables involving a Higgs and/or a weak boson, so that the last operator of
Eq.~\eqref{LCPV} is also irrelevant. The Wilson coefficient parameter space of
interest is therefore spanned by the $\{\cg, \ \cga,\ \chw,\ \chb,\ \ctw\}$
ensemble of free parameters.
In principle, the $\cg$ operator could be constrained by multijet processes, like
for the corresponding $CP$-even operator. However, this requires a dedicated
study, which is beyond the scope of this work.

In general, it is difficult to construct a new physics model that will
only induce $CP$-violating operators. On the other hand, the hypothesis of a
purely $CP$-odd Higgs boson is experimentally disfavored whereas the
experimental bounds on the Higgs boson being an admixture of $CP$-even and
$CP$-odd states are very weak~\cite{Khachatryan:2014kca,Khachatryan:2016tnr}.
Therefore, a more realistic setup would be a case where the Lagrangian contains
both $CP$-odd and $CP$-even operators. Deriving constraints on this new physics
configuration would then require a multidimenional fit of all $CP$-odd
and $CP$-even parameters. As a first study, we nevertheless consider the purely
$CP$-odd Lagrangian of Eq.~\eqref{LCPV} and leave the joint study of the impact
of both $CP$-even and $CP$-odd operators for future works.

The main effects that originate from the $\chb$
operator however arise from the Higgs coupling to the $Z$-boson, and can thus
always be reabsorbed by a redefinition of the $\chw$ operator,
\be
\chw \to \chw + t_{\sss W}^2 \, \chb \label{WtoZ}\ ,
\ee
where $t_{\sss W}=\tan \theta_{\sss W}$ is the tangent of the weak mixing angle
(as shown in the second Feynman rule of Figure~\ref{fig:cpvFR}).

\begin{table}
 \centering
  \renewcommand{\arraystretch}{1.6}
  \begin{tabular}{c|c|c|c|c|c}
    Process & $\cg$ & $\cga$ & $\chw$ & $\chb$ &$\ctw$ \\\hline\hline
    $p p \to h \to \gamma \gamma$ & $\star$ & $\star$ &  &  & \\\hline
    $p p \to h \to Z Z^{(*)} \to 4 \ell $
      & $\star$ & &  & $\star$ & \\\hline
    $p p \to h \to Z \gamma$ & $\star$ & & $\star$ & $\star$ & \\\hline
    $p p \to Z h \to \ell^+ \ell^- b \bar b$ & & & $\star$ & $\star$ & \\\hline
    $p p \to Z h \to \nu \bar \nu b \bar b$ & & & $\star$ & $\star$ &\\\hline
    $p p \to W h \to \ell \nu b \bar b$ & & & $\star$ & & \\\hline
    $p p \to h j j$  (VBF) & & & $\star$ & $\star$ & \\\hline
    $p p \to WW \to \ell \nu \ell' \nu'$ & & & $\star$ & & $\star$ \\
  \end{tabular}
  \renewcommand{\arraystretch}{1.0}
  \caption{List of LHC processes investigated in this work, presented together
    with their dependence, indicated by a star, on the EFT operators under
    consideration.}
   \label{tab1}
\end{table}

In order to probe the considered Wilson coefficient parameter space, we study a
set of processes that are particularly sensitive to CPV new physics effects in
the electroweak sector and that are shown in Table~\ref{tab1}, together with
their dependence on the different EFT parameters.
We consider simulations of collisions such as occurring at the LHC where the
hard process is calculated at the leading order accuracy and the fixed-order
result is then matched with parton
showers for a proper description of the QCD environment. Detector effects are
ignored, as well as next-to-leading order QCD corrections that could in
principle imply a dependence on the CPV triple-gluon operator ${\cal O}_{3G}$.

We can interpret the Lagrangian terms of Eq.~\eqref{LCPV} as the low-energy
manifestation of some new physics arising at a scale $\Lambda$, the details of
the ultraviolet completion being encoded in the $\tilde c$ coefficients.
Denoting by $g_{\rm NP}$ the strength of the new physics interactions, one can
derive
\be
  \frac{\tilde c}{\mW^2} \approx \frac{g_{\rm NP}^2}{\Lambda^2}\ .
\label{Lambdadef} \ee
This expression approximates the more precise relation that can be computed in
an ul\-tra\-vio\-let-complete setup, as shown for instance in the analyses of
Refs.~\cite{Henning:2014wua,Gorbahn:2015gxa,Drozd:2015kva}. In the next
sections, we adopt the choice of quoting our results in terms of the
dimensionless $\tilde c$ coefficients, but we also derive a more intuitive
estimation of the LHC sensitivity to new physics by extracting a bound on the
effective scale $\Lambda$ in the context of typical strongly-coupled
(so that $\Lambda>\Lambda_s$) and weakly-coupled (so that $\Lambda>\Lambda_w$)
scenarios. The
$\Lambda_s$ and $\Lambda_w$ limits are inferred from Eq.~\eqref{Lambdadef}, the
$g_{\rm NP}$ coupling being fixed to $4\pi$ and $g$ for the strongly-coupled and
weakly-coupled new physics cases respectively.
Deriving the $\Lambda_w$ and $\Lambda_s$ values enables us to verify whether the
phase space regions probed in our investigations of the CPV operators of
Eq.~\eqref{LCPV} are regions where the EFT approach is reliable. Our test is
based on a comparison of the hard scattering scale of the simulated collisions
with the $\Lambda_s$ and $\Lambda_w$ values, which differs from other methods
that have been proposed to assess the validity of the EFT approach~\cite{%
Degrande:2016dqg,Contino:2016jqw}. It should therefore be taken as a matter of
convention to translate limits on dimensionless $\tilde c$ coefficients to
limits on a mass scale. In particular, in theories where new physics effects are
only induced at the loop level, additional loop-suppression factors must be
incorportated.

\section{Connecting the effective approach to ultraviolet-complete models}
\label{sec:UV}

Although the EFT paradigm allows one to pursue a model-independent approach to
new physics, it is always important to reinterpret any EFT result in the
framework of specific ultraviolet-complete models. Maximizing the chances of
discovering new physics motivates to follow pragmatically both a top-down and a
bottom-up path. The explicit matching of an ultraviolet-complete theory to its
effective counterpart is however going beyond the scope of this work.

The simplest example incorporating an ultraviolet origin for the CPV new physics
operators of the effective Lagrangian of Eq.~\eqref{LCPV} consists of a setup
where the Standard Model is supplemented by new heavy fermions whose
interactions with
the Higgs boson feature explicit CPV effects. More precisely, we consider a set
of new heavy quarks,
\be
  \bigg\{ Q=\bpm T \\  B \epm\ , \quad T'\ , \quad B'\ \bigg\},
\ee
where $Q$ is a weak doublet of hypercharge 1/6, and where $T'$ and $B'$ are two
weak singlets of hypercharge 2/3 and -1/3 respectively. Yukawa interactions of
these new fields with the Higgs field $\Phi$ can be generically written as
\be\bsp
  {\cal L}_{\rm UV} = & \
       - y_B \, \bar Q  \Phi B'
       - i \tilde y_B \, \bar Q  \Phi \gamma_5 B'
       - y_T \, \bar Q  \cdot \Phi^\dag T' \\ &\ 
       - i \tilde y_T \, \bar Q  \cdot \Phi^\dag \gamma_5 T'
       + {\rm h.c.} \ ,
\esp \ee
where the dot product stands for the $SU(2)$-invariant scalar product and where
any possible mixing of the Standard Model quarks with the new heavy states is
neglected.
Such new fermions could appear, for example, in composite Higgs
models where fermionic partners to the third generation quarks are introduced to
trigger the breaking of the electroweak symmetry~\cite{Espinosa:2011eu}.

The integration out of the heavy fermions leads to the generation of several
effective $CP$ violating and $CP$ conserving operators. One obtains, for
instance, a non-vanishing dimension-six coupling of the Higgs field to the gluon
field strength tensor,
\be
  {\cal L}_{\rm EFT}= \frac{g_s^2}{16 \pi^2}
    \bigg[ \frac{\tilde y_B^2}{m^2_B} + \frac{\tilde y_T^2}{m^2_T}\bigg]
    \Phi^\dag \Phi \, G_{\mu\nu}^a {\widetilde G}^{\mu\nu}_a .
\ee
Mapping this operator to the Lagrangian of Eq.~\eqref{LCPV}, one gets the
matching condition
\be
  \tilde c_g = \frac{1}{16 \pi^2} \bigg[ \frac{\mW^2}{m^2_B} \tilde y_B^2 +
     \frac{\mW^2}{m^2_T} \tilde y_T^2 \bigg] \ ,
\ee
where the new physics coupling strength $g_{\rm NP}$ is identified with the CPV
Yukawa couplings, and where the new physics scale corresponds to the mass of the
heavy fermions.

The operators shown in Eq.~\eqref{LCPV} can also be generated in compositeness
models including composite scalars~\cite{Gripaios:2009pe,Sanz:2015sua}.
Depending on the vacuum structure~\cite{Galloway:2010bp}, the $CP$ symmetry can
be spontaneously broken and yield to CPV EFT operators
once the heavy scalars are integrated out~\cite{Croon:2015naa,%
No:2015bsn}.

On different grounds, many popular extensions of the Standard Model contain an
extended Higgs sector that includes, \textit{e.g.}, new scalar weak singlets or
doublets. Explicit CPV in the Higgs sector does not however induce effective
operators such as those shown in the Lagrangian of Eq.~\eqref{LCPV}, but instead
modifies the magnitude of the Standard Model Higgs couplings~\cite{%
Gorbahn:2015gxa}. Most beyond the Standard Model theories nonetheless generally
exhibit a particle spectrum with many new degrees of freedom, whose integration
out in contrast leads to new Lorentz structures in the interactions of the
Standard Model fields~\cite{Carena:2000ks,Carena:2001fw,Carena:2002bb,%
Ellis:2007kb,Carena:2015uoe}.

\section{LHC Run~I bounds on CPV EFT operators}~\label{sec:Run1}

Constraints on the Wilson coefficients appearing in the Lagrangian of
Eq.~\eqref{LCPV} can be obtained by analyzing Higgs boson and vector boson decay
and production rates once predictions in the EFT framework are compared with LHC
Run~I measurements. The most stringent Run~I constraints on the $\cg$ and $\cga$
coefficients arise from the results of the CMS and ATLAS combination for Higgs
boson production and decay in the $g g \to h \to \gamma\gamma$
channel~\cite{Khachatryan:2016vau}, the associated signal strength being given
by
\be
  \mu_{\rm LHC}^{gg\to h\to\gamma\gamma} =  1.09_{-0.10}^{+0.11}  \ .
\label{eq:signalstrength}\ee
While other limits on the new physics contributions to the Higgs boson couplings
to gluons and photons are available, these are extracted under the assumption
that either the Higgs boson width or its production rate is the Standard Model
one. We thus restrict
ourselves to the use of Eq.~\eqref{eq:signalstrength}. The corresponding
theoretical predictions (see Appendix~\ref{appA} for technical details on the
simulations performed in this work) can be fitted by a quadratic function of the
CPV $\cg$ and $\cga$ parameters,
\be
  \mu_{\rm EFT}^{g g \to h \to \gamma\gamma} = 1.0
   + 2.0 \times10^{7} \cga^2
   - 1.3 \times 10^3 \cga \cg
   + 2.0 \times 10^5 \cg^2 \ ,
\ee
where the absence of linear terms stems from the vanishing interferences between
the new physics and the Standard Model contributions.

\begin{figure}
  \centering
  \includegraphics[width=0.83\columnwidth]{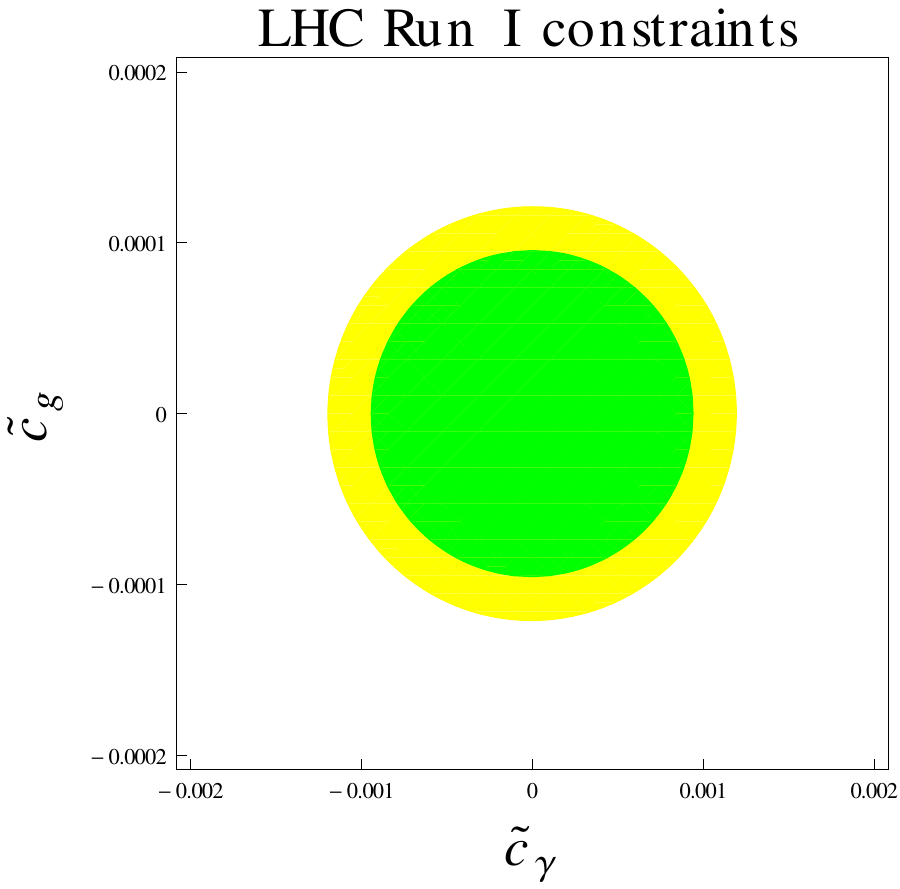}
  \includegraphics[width=0.79\columnwidth]{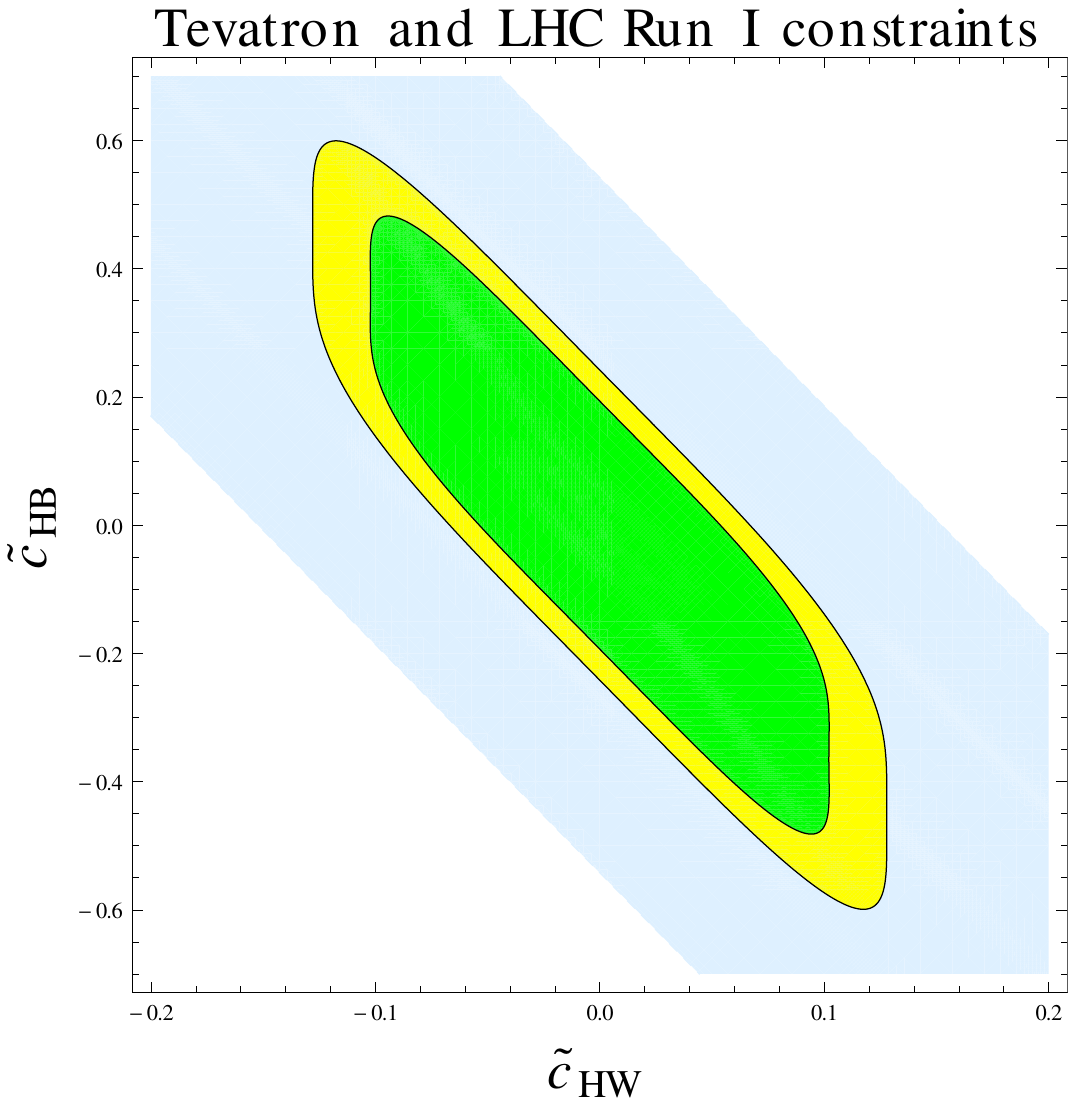}
  \caption{Collider bounds on several of the effective operators considered in
    the Lagrangian of Eq.~\eqref{LCPV}. We show parameter space regions in
    agreement with LHC Run I data in the $(\cga, \cg)$ (left) and $(\chw, \chb)$
    (right) plane at the $1\sigma$ (green) and $2\sigma$ (yellow) level, and
    the region allowed by Tevatron data at the 95\% confidence level is
    indicated by the blue area.}
  \label{fig:Run1cgcga}
\end{figure}

On the other hand,
electroweak Higgs boson production processes allow to constrain both the $\chw$
and $\chb$ coefficients on the basis of LHC Run~I and Tevatron data. Starting
with Higgsstrahlung ($VH$) signal strengths, the CVP EFT framework depicted by
Eq.~\eqref{LCPV} leads to theoretical predictions that can be fitted
quadratically by
\be\bsp
  \mu_{\rm EFT}^{ZH,{\rm ~LHC}}=&\ 1.0+145.6 \big(\chw+t_W^2\chb\big)^2\ ,\\
  \mu_{\rm EFT}^{WH,{\rm ~LHC}}=&\ 1.0+52.3\ \chw^2\ , \\
  \mu_{\rm EFT}^{ZH,{\rm ~Tev}}=&\ 1.0+104.7 \big(\chw+t_W^2 \chb\big)^2\ ,\\
  \mu_{\rm EFT}^{WH,{\rm ~Tev}}=&\ 1.0 +  35.12\ \chw^2  \ ,
\esp\label{fitVH}\ee
for the LHC and the Tevatron colliders respectively. These must be compared with
the corresponding measurements~\cite{Khachatryan:2016vau},
\be\bsp
  \mu_{\rm LHC}^{WH} =&\  0.88^{+0.40}_{-0.38}\ , \\
  \mu_{\rm LHC}^{ZH} =&\  0.80^{+0.39}_{-0.36}\ , \\
  \mu_{\rm Tev}^{VH} =&\  1.59^{+0.69}_{-0.72}\ ,
\esp\ee
the Tevatron value being mainly driven by the $ZH$ production mode with a final
state signature containing either zero or two leptons~\cite{Ellis:2012xd,%
Ellis:2013ywa}. Additional constraints can be induced by vector-boson fusion
(VBF) Higgs production results, and in particular by the $WW$ channel (WBF) that
contributes to the signal strength result with a weight of 80\%. A fit of the
EFT theoretical predictions gives
\be
  \mu^{\rm WBF,~LHC}_{\rm EFT} = 1.0 + 25.3 \, \chw^2\ ,
\label{fitVBF}\ee
which can be confronted to the Run~I results,
\be
  \mu_{\rm LHC}^{\rm WBF} = 1.18^{+0.25}_{-0.23}\ .
\ee
Although VBF data is more precise and features smaller error bars than in the
$VH$ case, the
sensitivity of the $VH$ production processes to the CPV EFT operators is then
expected to be higher than in the VBF case, as pointed out by the numerical
factors multiplying the $\tilde c$ terms found in Eq.~\eqref{fitVH} and
Eq.~\eqref{fitVBF}.

\begin{table}
  \centering
  \renewcommand{\arraystretch}{1.6}
  \setlength\columnsep{20pt}
  \begin{tabular}{c||c||c|c}
    Coefficient & Limit  & $\Lambda_s$  & $\Lambda_w$ \\
    \hline\hline
    $|\cg|$  & 1.2 $\times 10^{-4}$ & 92~TeV & 4.4~TeV\\\hline
    $|\cga|$ & 1.2 $\times 10^{-3}$ & 29~TeV & 1.4~TeV\\\hline
    $|\chw|$ & 0.06 & 4.1~TeV & [0.2~TeV] \\\hline
    $|\chb|$ & 0.23 & 2.1~TeV & [0.1~TeV] \\\hline
    $|\ctw|$ & 0.18 & 2.4~TeV & [0.1~TeV] \\
  \end{tabular}
  \caption{LHC Run~I constraints on the Wilson coefficients associated with the
    CPV EFT operators given in Eq.~\eqref{LCPV} (second column), also casted
    under the form of a bound on
    the effective scale for strongly-coupled (third column)
    and weakly-coupled (fourth column) new physics. The brackets indicate that
    the limit has been extracted under conditions not compatible with the
    expected EFT range of validity. We refer to Section~\ref{sec:EFT} for
    details on how the limits on $\Lambda_{w,s}$ are defined.}
   \label{tab:Run1}
  \renewcommand{\arraystretch}{1.0}
\end{table}

From the relations derived above, we perform a $\chi^2$ fit of LHC data and
extract limits on the effective parameters. The results are shown in
Table~\ref{tab:Run1}, as well as in Figure~\ref{fig:Run1cgcga} where we have
projected them in the $(\cga, \cg)$ (left) and $(\chw, \chb)$ (right) planes.
Our procedure relies on neglecting the $WH$ Tevatron information and on
averaging the experimental errors. We observe that operators which affect
processes that are loop-suppressed in the Standard Model are more strongly
constrained, the maximum allowed value for the associated $\cg$ and $\cga$
parameters being of the order of 0.001 for an effective scale being the
$W$-boson mass. Equivalently, this corresponds to probing an effective scale
reaching
the multi-TeV regime for typical strongly-coupled or weakly-coupled new physics.
In contrast, current limits on the electroweak operators and the corresponding
$\chw$, $\chb$ and $\ctw$ parameters must be carefully interpreted in the
case of weakly-coupled new physics. The corresponding bound on the effective
scale indeed implies that this scale may be too small to guarantee the validity
of the EFT all over the limit extraction procedure. The results finally
also depict the
strengthening of the Tevatron constraints once LHC Run~I measurements are
accounted for.

The $\chw$, $\chb$ and $\ctw$ are hence currently only loosely constrained by
data. In the rest of this work, we demonstrate how future LHC data at a higher
center-of-mass energy is expected to provide better handles on the
associated operators, and we design novel ways to use the 13 TeV future results
to enhance the corresponding LHC sensitivity.

In addition to the processes introduced above, the $\chw$ and $\chb$ parameters
could also be constrained by investigating Higgs boson production and decay into
a
four-leptonic final state. Fitting the theoretical predictions, the related LHC
signal strength is given, in the CPV EFT context, by
\be
  \mu_{\rm EFT}^{pp\to h\to 4\ell,~{\rm LHC}}
   = 1.0 + 123.3 \big(\chw + t_W^2 \, \chb \big)^2\ ,
\ee
that we can compare the ATLAS and CMS combined value~\cite{Khachatryan:2016vau}
of
\be
  \mu_{\rm LHC}^{pp\to h\to 4\ell} = 1.13^{+0.34}_{-0.31} \ .
\ee
This process is also strongly affected by the $\cg$ parameter, so that
meaningful constraints should be extracted from a multidimensional fit. However,
we have verified that the predictions barely depend on this
higher-dimensional coupling once its range is restricted by the current
constraints. We therefore neglect it in the subsequent analysis.

Table~\ref{tab:Run1} finally also includes a bound on the $\ctw$ coefficient
that we have extracted from the LHC Run~I $W$-boson pair
production cross section measurement~\cite{Aaboud:2016mrt},
\be
  \sigma_{WW}= 71.1 \pm 1.1 ({\rm stat})^{+5.7}_{-5.0} ({\rm syst})
     \pm 1.4 ({\rm lumi})~{\rm pb}\ .
\ee
Making use of the Standard Model predictions computed at the next-to-next-to-%
leading order accuracy in QCD~\cite{Campbell:2011bn,Heinemeyer:2013tqa,%
Gehrmann:2014fva},
\be
  \sigma_{WW}^{({\rm NNLO})} = 63.2^{+1.6}_{-1.4} ({\rm scale}) \pm 1.2
     ({\rm PDF})~{\rm pb} \ ,
\ee
we can derive a signal strength value $\mu_{\rm LHC}^{WW}$ by computing the
largest possible allowed deviation in the ratio of data to theory once all
errors are added in quadrature~\cite{Aad:2016wpd},
\be
  \mu_{\rm LHC}^{WW} = 1.13 \pm 0.07 \ .
\ee
This result can then be confronted to the CPV EFT fitted signal strength
\be
  \mu_{\rm EFT}^{WW} = 1.0 +8.0 \, \ctw^2 \ .
\ee
Additional constraints could also in principle be derived from $WZ$ and $ZZ$
total cross section measurements, but these are found less sensitive to the
considered new physics operators, and are thus ignored.

Experimental collaborations have also performed specific studies on anomalous
Higgs couplings to the Standard Model vector bosons in the dilepton and the
four-lepton channel~\cite{Khachatryan:2014kca,Khachatryan:2016tnr}. The general
line of these analyses relies on Higgs-boson production via gluon fusion, with a
subsequent decay of the Higgs boson into a pair of vector bosons. The two
channels that have been considered are the $h\to W^{+}W^{-}\to 2\ell 2\nu$ and
$h\to ZZ\to 4\ell$ ones, and the analysis strategy involves several kinematic
discriminants being several invariant masses. The results are presented in terms
of an effective fractional cross section which describes the allowed amount of
deviation with respect to the Standard Model expectation. The Run~I results have
been found not conclusive due to a too low statistics, and the 13~TeV results
still allow for a large amount of $CP$-violation.

\section{Prospective LHC studies on the basis of inclusive measurements}
\label{sec:future}

In this section, we evaluate the LHC sensitivity to new physics effects modelled
by the effective operators of the Lagrangian of Eq.~\eqref{LCPV}, assuming an
integrated luminosity of either 300~fb$^{-1}$ (to be achieved by 2020) or
3000~fb$^{-1}$ (the goal of the High-Luminosity LHC program). The estimate of
the prospects for the precise determination of the Higgs couplings has been
deeply studied by all experimental collaborations, and the ATLAS collaboration
has in particular presented results including a channel breakdown~\cite{%
ATL-PHYS-PUB-2014-016}. The pieces of information relevant for our study are
summarized in Table~\ref{tab:prospects} under the form of the expected precision
on the signal strengths corresponding to various Higgs-boson production and
decay subprocesses, the theory errors being omitted for brevity.

\begin{table}
  \renewcommand{\arraystretch}{1.6}
  \center
  \begin{tabular}{l| c | c l c |l}
   Channel & $\Delta \mu/\mu$ - 300~fb$^{-1}$
      & $\Delta \mu/\mu$ - 3 ab$^{-1}$\\ \hline \hline
   $h \to \gamma \gamma$ (jet veto) & 0.13 (0.09) & 0.09 (0.04) \\ \hline
   $h \to ZZ$ (gluon fusion) & 0.12 (0.07) & 0.11 (0.04) \\
   $h \to WW$ (jet veto) & 0.18 (0.09) & 0.16 (0.05) \\
   $h \to \gamma \gamma$ (VBF) & 0.47 (0.43) & 0.22 (0.15) \\
   $h \to \gamma \gamma$ ($WH$) & 0.48 (0.48) & 0.19 (0.17) \\ \hline
   $h \to Z Z$ ($VH$) & 0.35 (0.34) & 0.13 (0.12) \\
   $h \to Z Z$ (VBF) & 0.36 (0.33) & 0.21 (0.16) \\
   $h \to WW$ (VBF) & 0.21 (0.20) & 0.15 (0.09) \\ 
   $h \to b \bar b$ ($ZH$) & 0.29 (0.29) & 0.14 (0.13) \\ 
   $h \to b \bar b$ ($WH$) & 0.57 (0.56) & 0.37 (0.36) \\ 
\end{tabular}
  \caption{Expected accuracy on the Higgs signal strength measurements for
    different luminosities and different channels, as extracted from
    Ref.~\cite{ATL-PHYS-PUB-2014-016}. From a study of the $pp\to h \to
    \gamma\gamma$ process (first block of the table), one can extract
    constraints on the $\cga$ and $\cg$ parameters. The next four channels
    (second block of the table) provide information on the $\cga$, $\cg$, $\chb$
    and $\chw$ Wilson coefficients while all other processes (last block of the
    table) probe the $\chb$ and $\chw$ parameters.}
  \label{tab:prospects}
  \renewcommand{\arraystretch}{1.0}
\end{table}

The information embedded in the table allows for a global fit of all the Wilson
coefficients included in the Lagrangian of Eq.~\eqref{LCPV}. The three sets of
processes under consideration (separated by horizontal lines in the table) can
however be used to set bounds on independent pairs of operators, which motivates
the simpler procedure adopted in the following. For instance, a precise
measurement of the Higgs-boson properties in the $pp\to h\to\gamma\gamma$
channel, that is dominated by gluon-fusion production, would provide information
on the pair of $\cga$ and $\cg$ parameters whereas investigations of VBF or $VH$
Higgs-boson production events where the Higgs boson decays into a weak-boson
pair or a $b\bar b$ pair yield independent information on the $\chb$ and $\chw$
parameters. As a consequence, we focus on two-dimensional fits that are
also easier to represent.

\begin{figure}
  \centering
  \includegraphics[width=0.79\columnwidth]{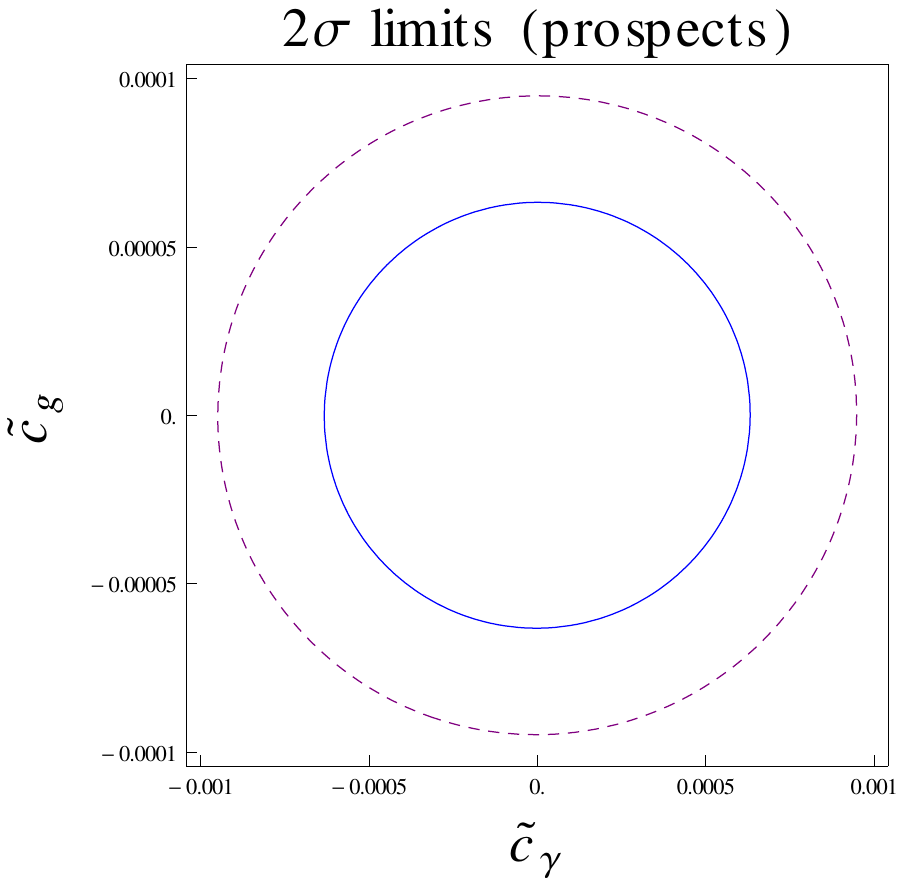}
  \includegraphics[width=0.75\columnwidth]{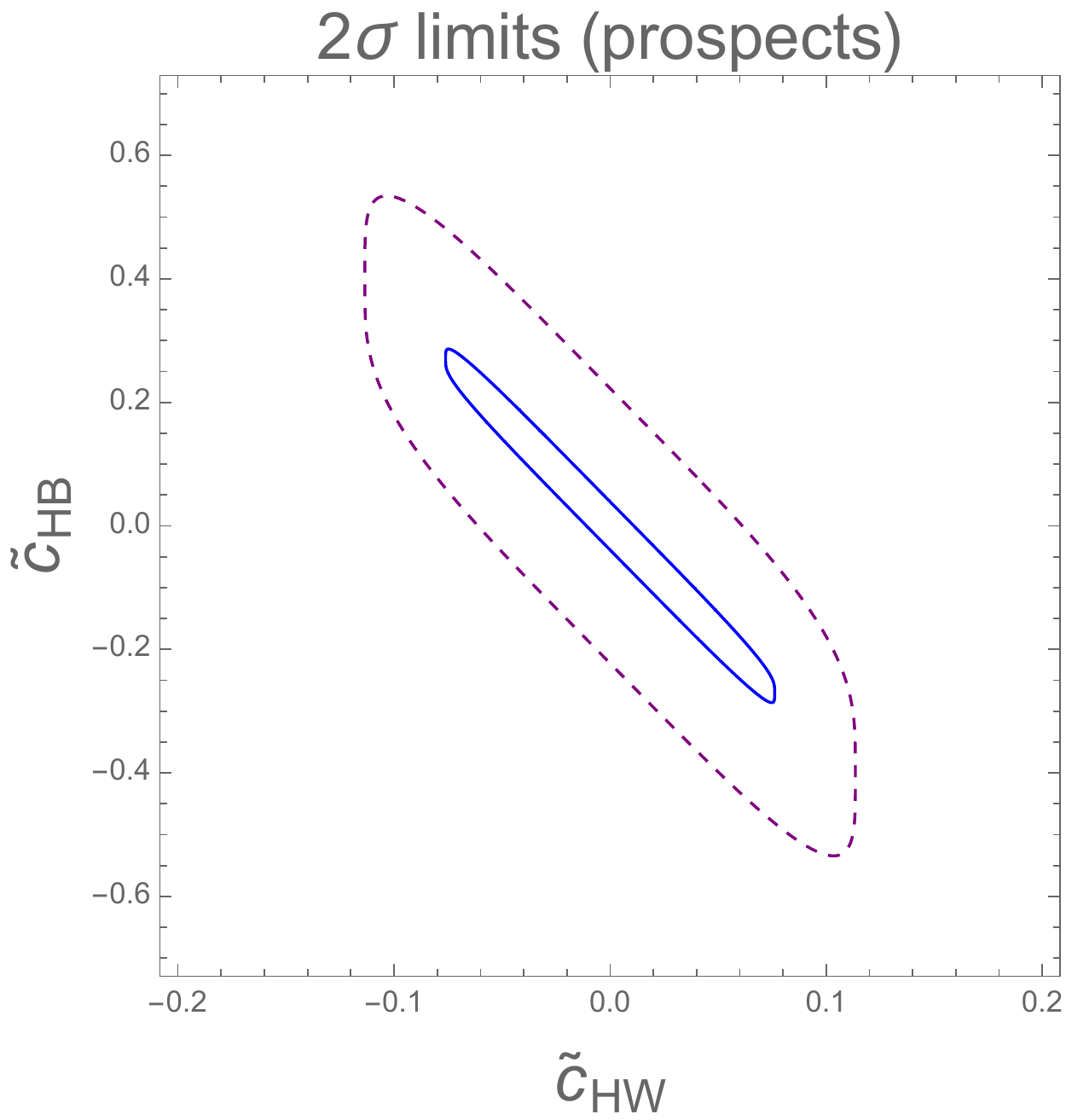}
  \caption{LHC sensitivity to the $\cga$ and $\cg$ (left) and on the $\chw$ and
    $\chb$ parameters (right). We show the 95\% confidence level reach for an
    integrated luminosity of 300~fb$^{-1}$ (dashed purple) and 3000~fb$^{-1}$
    (solid blue), neglecting the effects of the theoretical uncertainties.}
  \label{fig:cgcga}
\end{figure}

Theoretical predictions for the signal strength associated with the $g g\to h\to
\gamma\gamma$ channel are given, in terms of the $\cg$ and $\cga$ parameters, by
the quadratic fitting function
\be
  \mu_{\rm EFT}^{gg\to h\to\gamma\gamma} =
     1.0 + 2.0 \times 10^5 \, \tilde{c}_{\gamma}^2
   - 1.5 \times 10^4 \, \tilde{c}_{\gamma} \tilde{c}_{g}
   + 2.0 \times 10^7 \, \tilde{c}_{g}^2\ ,
\ee
once a basic selection is applied on the signal. Confronting those predictions
to the expectations presented in Table~\ref{tab:prospects} thus allows to
extract the LHC sensitivity to the $\cg$ and
$\cga$ Wilson coefficients. We show results in the left panel of
Figure~\ref{fig:cgcga} for a luminosity of 300~fb$^{-1}$ (dashed purple) and
3000~fb$^{-1}$ (solid blue) of proton-proton collisions at a center-of-mass
energy of 13 TeV.

Similarly, we can
extract bounds on the remaining coefficients by focusing on processes
independent of the $\cga$ and $\cg$ parameters like those presented in the last
panel of Table~\ref{tab:prospects}. The predictions for the three most relevant
signal strengths are given by
\be\bsp
\mu_{\rm EFT}^{pp\to ZH} =& 1.0 + 168 \,  (\chw+  t_W^2 \, \chb)^2\ , \\
\mu_{\rm EFT}^{pp\to WH} =& 1.0 + 53 \,  \chw^2\ , \\
\mu_{\rm EFT}^{\rm WBF}  =& 1.0+ 38 \, \chw^2 \ .
\esp\ee

Besides the channels described above, measurements related to the rare
$h\to Z \gamma$ decay also allow for the extraction of constraints on the $\chw$
and $\chb$ parameters, as the corresponding signal strength is sensitive to
these two EFT operator coefficients,
\be 
  \mu_{\rm EFT}^{h\to Z\gamma} = 1+ 6100 \, (\chw+t_w^2 \chb)^2 \ .
\ee
The prospects on limit setting by studying this rare Higgs boson decay mode have
been evaluated for 3000~fb$^{-1}$ of LHC collisions~\cite{ATL-PUB-2014-006},
\be
  \mu_{\rm LHC}^{h\to Z\gamma} =
     1.00^{+0.25}_{-0.26}~{\rm (stat.)}~{}^{+0.17}_{-0.15}~{\rm (syst.)} \ ,
\ee
so that the predictions can be compared to the experimental expected value.

The resulting constraints on the $\chb$ and $\chw$ parameters are shown on the
right panel of Figure~\ref{fig:cgcga}, when all the channels described above are
accounted for.

On different grounds, the $\ctw$ coefficient can be constrained as indicated in
Section~\ref{sec:Run1}, on the basis of $W$-boson pair production total rates.
Predictions for the corresponding signal strength read,
\be
  \mu_{\rm EFT}^{WW} = 1.0 + 9.3 \, \ctw^2\  .
\ee
The precision on the related experimental expectation
is however tightly bound both to experimental effects and to the accuracy
of the theoretical predictions that is currently the next-to-next-to-%
leading order in QCD~\cite{Grazzini:2016ctr}. We can optimistically
estimate the total error to be of the order of 5\%, which would lead to a
moderate enhancement of the expected constraints on $\ctw$ by a
factor of about 2 with respect to the results of Table~\ref{tab:Run1}.

Comparing the Run~I results (Figure~\ref{fig:Run1cgcga}) with the
High-Luminosity LHC prospects (Figure~\ref{fig:cgcga}), we observe that an
improvement of a factor of about 2 can be expected. While
this mild strengthening of the constraints implies that the EFT is still used in
a range where it is valid, this also shows that the current bounds will not
drastically change during the next 20 years when solely signal strengths are
used. In the next section, we will show
how a more dramatic improvement could be achieved by making use of differential
distributions. For specific channels like the $VH$ or the diboson ones,
differential information is actually expected to be more powerful than what
could be obtained from total rate
measurements~\cite{Ellis:2012mj,Ellis:2014jta,Ellis:2014dva,Butter:2016cvz}.

\section{Prospective LHC studies using differential information}
\label{sec:futurediffs}

Derivative EFT operators have a momentum dependence, illustrated in the Feynman
rules of Figure~\ref{fig:cpvFR}, that could be exploited by focusing on phase
space regions where the momentum transfer is large.
As the $\cg$ and $\cga$ Wilson coefficients are already
well cornered by total rate measurements in the Higgs boson dominant production
(gluon-fusion) mode once a decay into photons is accounted for, we move on with
the use of differential distributions to design an analysis allowing one
to improve the expectation on the $\chw$, $\chb$ and $\ctw$ parameters. These
are all currently relatively less constrained by total rates, and the future
prospects have not been found very exciting.

A complication may arise from the fact that in general, as stated in
Section~\ref{sec:EFT}, the EFT Lagrangian stemming from an ultraviolet-complete
theory contains both $CP$-even and $CP$-odd operators. One must thus in 
principle construct observables that genuinely capture the CPV effects.
Some extensive studies along these lines have been conducted in previous
works~\cite{Han:2009ra,Christensen:2010pf}, where key observables are designed
on
the basis of triple products of momenta. This has been shown to be sensitive to
the interactions of the Higgs boson with a pair of weak gauge bosons.
On different lines, the EFT derived from many ultraviolet-complete models, like
supersymmetry or the Two-Higgs-Doublet Model, features effective couplings of
the Higgs boson to a gauge-boson pair whose CPV component is loop-suppressed. As
a consequence, the CPV contributions to cross sections, that are also the
quantities usually constrained by previous experimental searches, are always
small.
Exceptions exist for cases where there is a large admixture of $CP$-odd and
$CP$-even states that can be degenerate, and/or when the theory exhibits large
$CP$-violating phases~\cite{Fuchs:2017wkq,Fowler:2009ay}.
Earlier studies have also attempted to construct angular variables that directly
probe the interferences between the $CP$-odd, $CP$-even and the Standard Model
contributions in the VBF production mode~\cite{Hankele:2006ma} as well as those
induced by the coupling of the Higgs boson to a pair of $Z$-bosons~\cite{Belyaev:2015xwa}.

Another option to get sensitivity to $CP$-violation effects may rely on the
usage of phase-space correlations, which may become feasible as more data is
being recorded by the experiments. A variety of decay modes could be
considered~\cite{Delaunay:2013npa}.
For instance, the diphoton $h\to \gamma\gamma$ channel could be promising
provided that the photon polarization, a quantity directly related to $CP$
violation, could be measured. This can be achieved through the study of the
opening azimuthal angle between the two photons, that is expected to be in the
$[10^{-4}-10^{-3}]$ range and that thus lies at the resolution limit of the
ATLAS and CMS pixel detectors. It is thus possible to observe substantial
effects in parts of the phase space by choosing suitable cuts, but this is
unrealistic at the moment as the LHC integrated luminosity is still limited.
Another example concerns $Wh$ production, but this requires to be able to
separate the different initial-state helicity combinations. This can be 
performed through severe selections necessary as the $q\bar q$ initial-state is
symmetric in the context of a $pp$ collision.

\begin{figure}
  \centering
  \includegraphics[scale = 0.11]{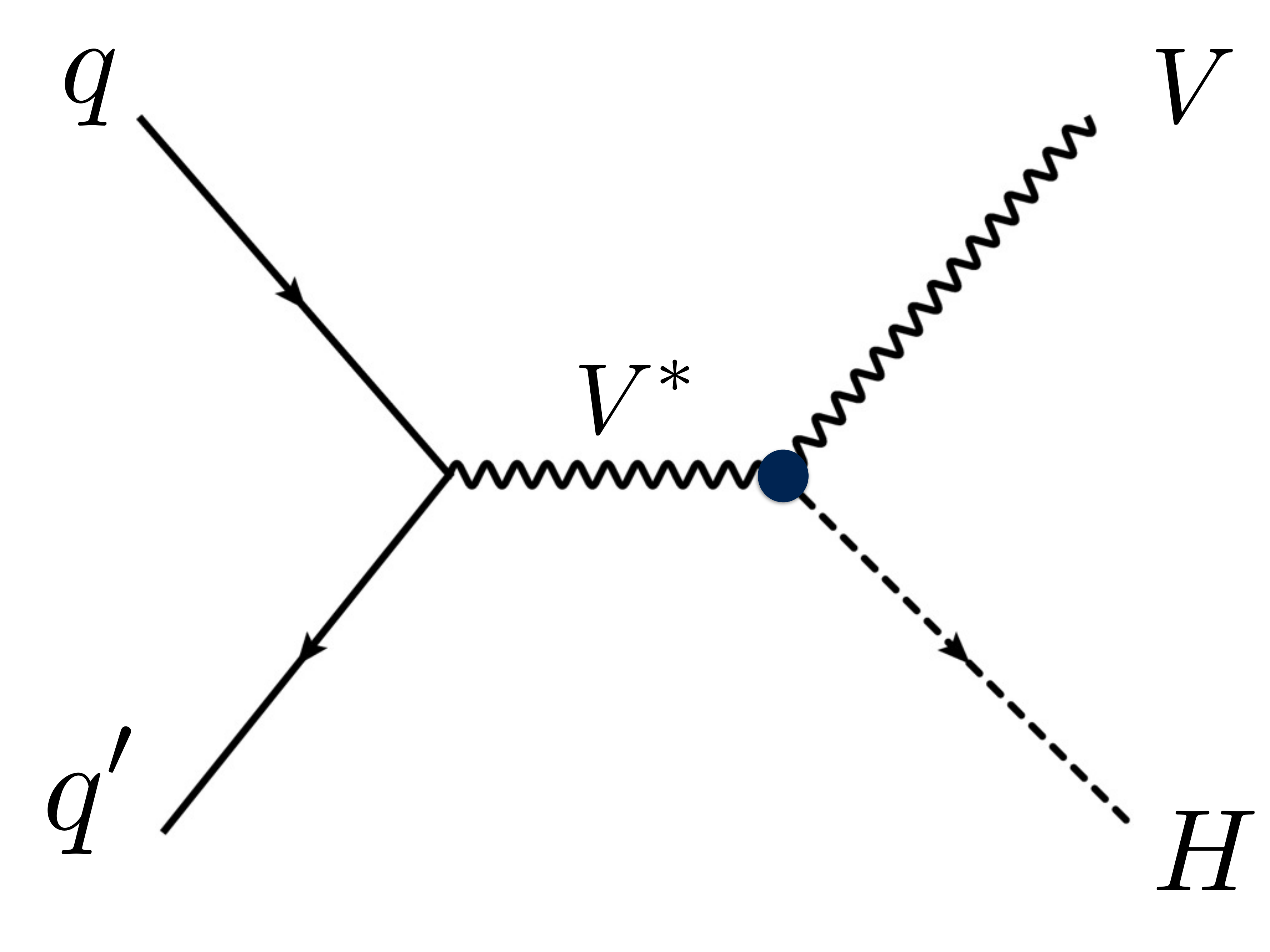} \hspace{0.2cm}
  \includegraphics[scale = 0.11]{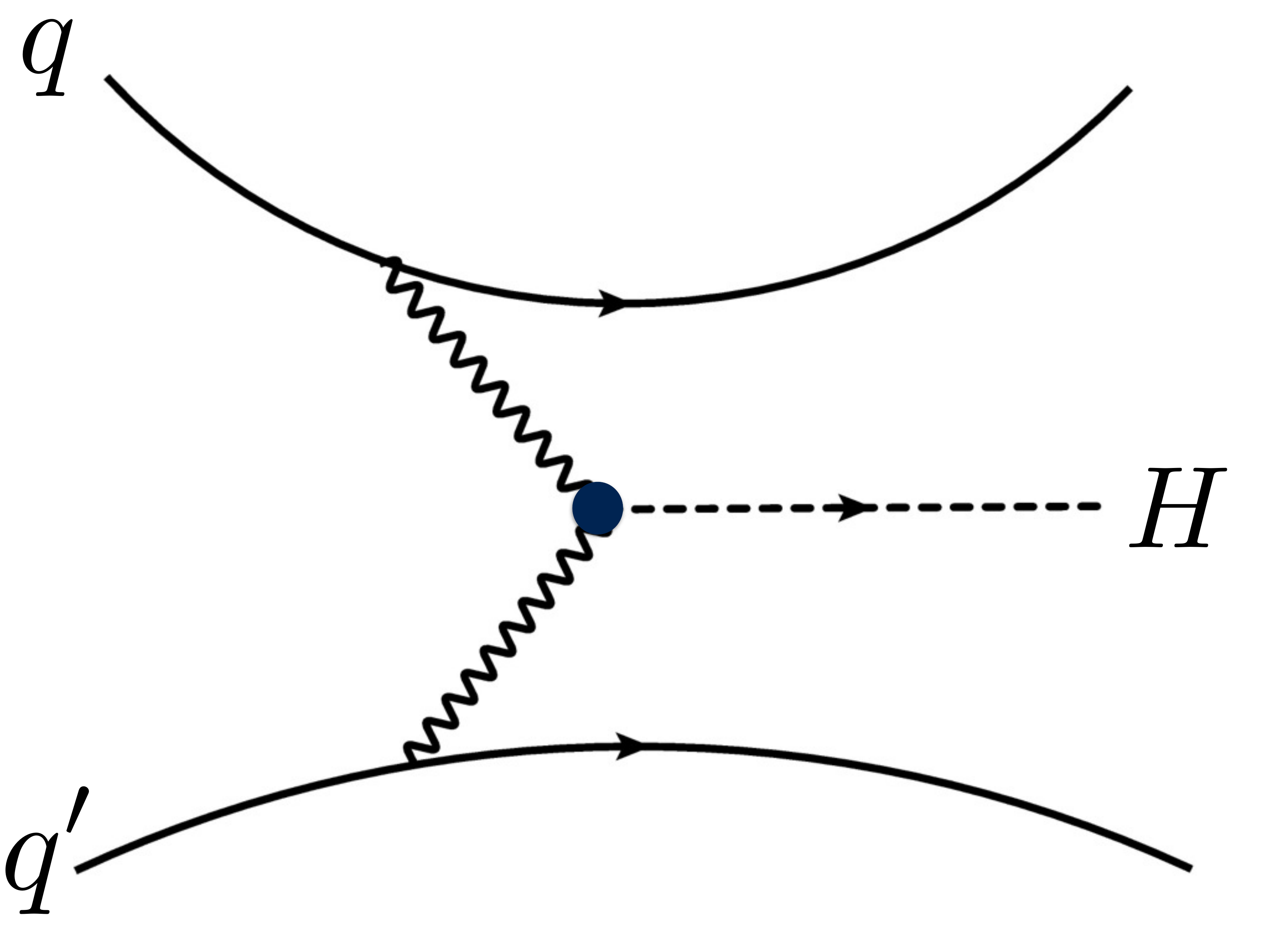}
  \includegraphics[scale = 0.11]{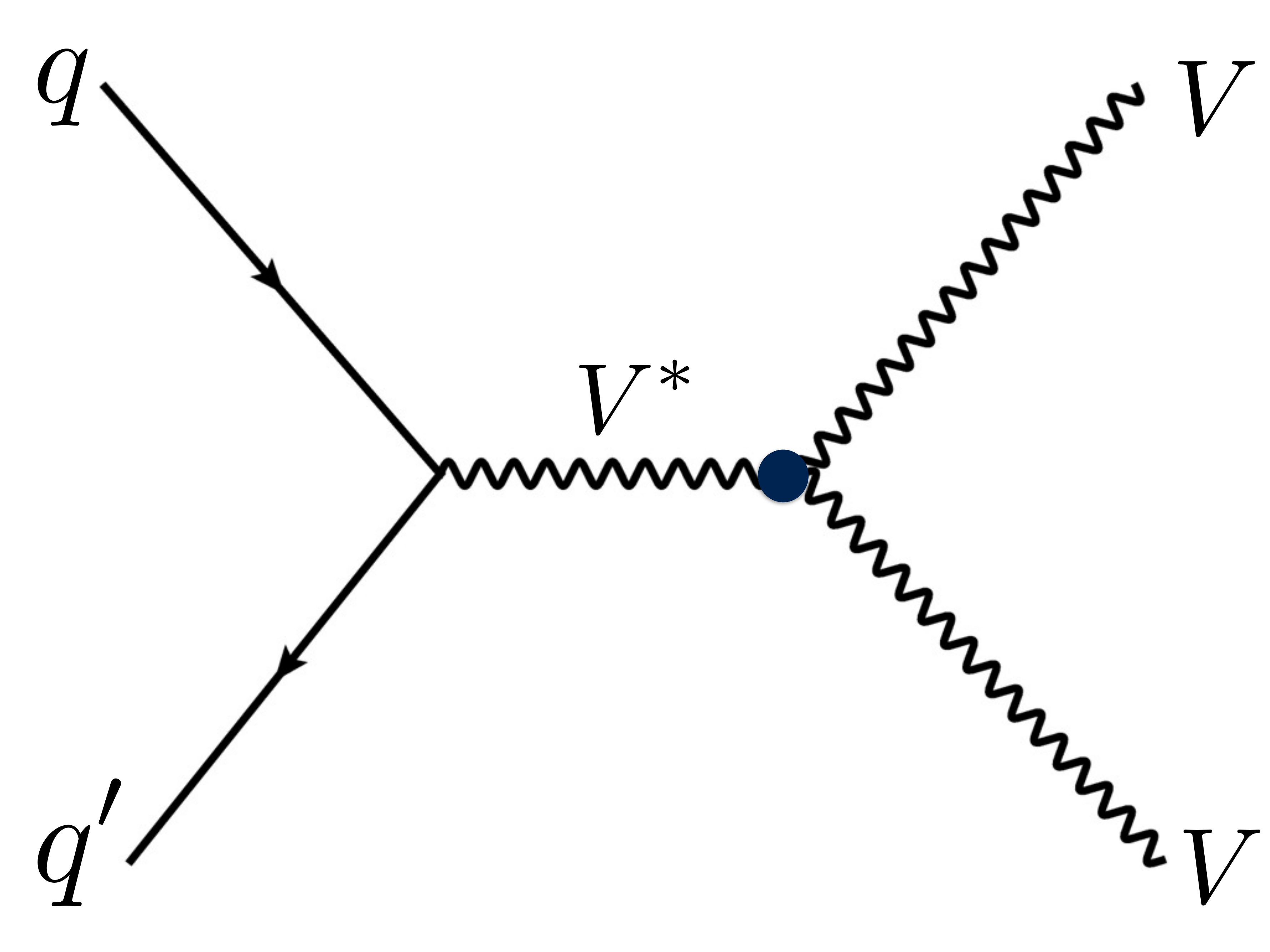}
  \caption{Representative Feynman diagrams for the considered Higgs and weak
    boson production mechanisms, namely for $VH$ associated production (left),
    VBF Higgs boson production (center) and diboson production (right).}
  \label{fig:channels}
\end{figure}

To study these momentum-dependent couplings in LHC collisions at a
center-of-mass energy of 13~TeV, we consider the electroweak processes shown in
Figure~\ref{fig:channels}, where Higgs and/or weak bosons are produced possibly
in association with jets. More precisely, we investigate the associated
production of a Higgs and a weak boson ($VH$), Higgs-boson production by vector
boson fusion and diboson production ($VV$). Concerning the boson decays, we
consider both the four-lepton mode traditionally studied for
$CP$-violation analyses~\cite{Choi:2002jk,Godbole:2007cn,Hagiwara:2009wt,%
Englert:2010ud} and novel channels, the seeds for some of them having been
introduced in earlier works~\cite{Plehn:2001nj,Desai:2011yj,Stolarski:2012ps,%
Freitas:2012kw,Chen:2013ejz,Bishara:2013vya,Chen:2014pia,Chen:2014ona}.

Technical details on the LHC collision simulations that we have performed are
given in Appendix~\ref{appA}.

\subsection{$VH$ Higgs and weak boson associated production}

In the following, we focus on the associated production of a Higgs and a weak
boson when the weak boson decays into either a single-lepton or a dilepton final
state. The Higgs boson is additionally considered to decay into a final-state
system from which it could be fully reconstructed, the precise definition of
this system being therefore not relevant.

\begin{figure}
  \centering
  \includegraphics[width=0.95\columnwidth]{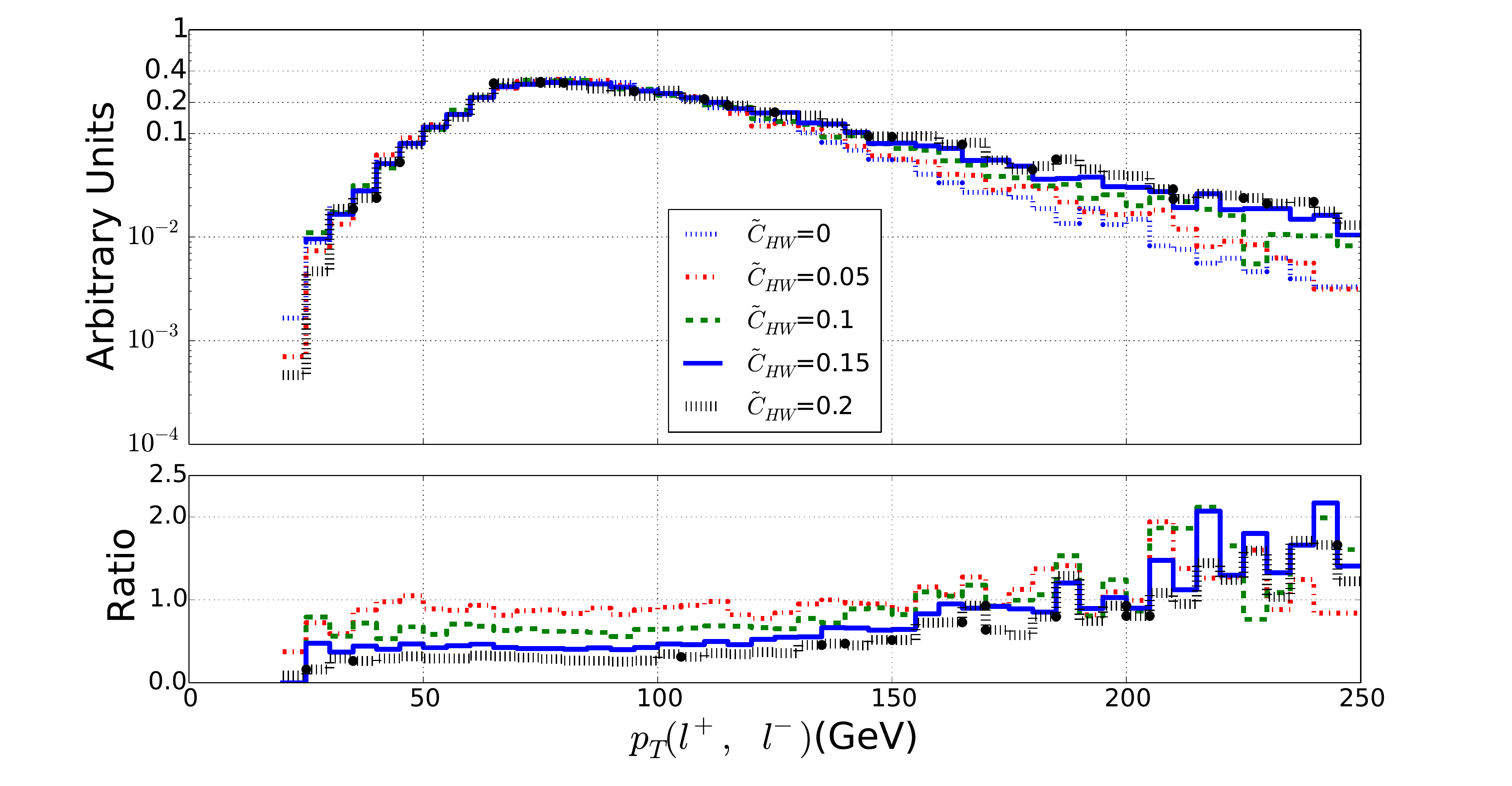}\\
  \includegraphics[width=0.95\columnwidth]{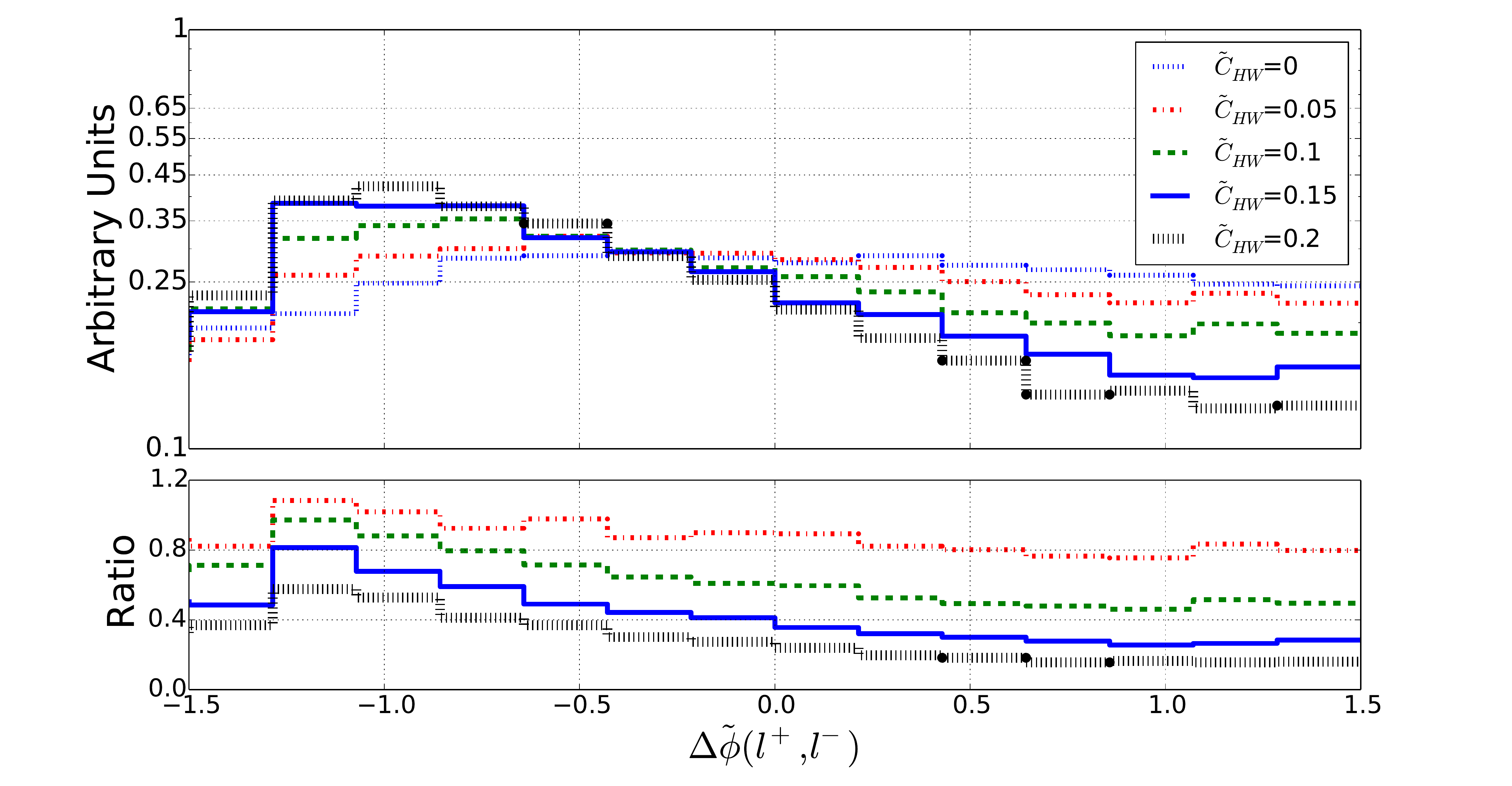}
  \caption{Representative kinematical properties of a dilepton system issued
  from the decay of a $Z$-boson when the latter is produced in association with
  a Higgs boson in LHC collisions at a center-of-mass energy of 13~TeV. We
  consider the scalar sum of the transverse momenta of the two
  leptons (top) and their angular separation in azimuth (bottom). We
  allow for different values for the $\chw$ parameter and we present, in the
  lower panels, the bin-by-bin ratio of the new physics predictions to the
  Standard Model expectation.}
  \label{fig:ZH}
\end{figure}

When the Higgs boson is produced together with a leptonic
$Z$-boson, we can make use of the kinematical properties of the two final-state
leptons to get handles on any possible EFT deviation. This is illustrated by the
two distributions shown in Figure~\ref{fig:ZH}, namely the scalar sum of the
transverse momenta of the two leptons $\ell^+$ and $\ell^-$ (upper panel),
\be
  p_T(\ell^+, \ell^-) = p_T({\ell^+}) + p_T({\ell^-})\ ,
\ee
and their angular separation in azimuth (lower panel) defined by
\be
  \Delta\tilde\phi(\ell^+,\ell^-) = |\Delta\phi(\ell^+,\ell^-)|-\frac{\pi}{2}\ .
\label{dphitilde}\ee

In the Standard Model, the $p_T(\ell^+, \ell^-)$ distribution exhibits first a
peak for
$p_T(\ell^+, \ell^-)\sim60$~GeV before it slowly falls down for larger values.
We then allow for a positive non-vanishing $\chw$ parameter varying in the range
$[0, 0.2]$. Although this extends the range allowed by the current constraints
when EFT operators are considered one-by-one (see Table~\ref{tab:Run1}), this
conservatively accounts for potentially weaker constraints that could stem
from a  EFT
fit. We observe that the EFT effects tame the decrease of the distribution for
large $p_T(\ell^+, \ell^-)$ values, as a result of the enhanced EFT impact when
the momentum transfer is large. Deviations of a factor of up to two are found,
while one still lies within the EFT range of validity. Other EFT operators
could also affect the predictions, like the ${\cal O}_{HW}$ and ${\cal O}_{HB}$
operators of the Lagrangian of Eq.~\eqref{LCPV}, and the obtained behaviour turns
out to be similar. This suggests to define, as a handle for characterizing new
physics, the efficiency $\varepsilon(\tilde c, p_T^{\rm cut})$ that depends on
the Wilson coefficient $\tilde c$ and on a minimum value $p_T^{\rm cut}$ for the
$p_T(\ell^+, \ell^-)$ observable,
\be
  \varepsilon({\tilde c}) = \frac{1}{\sigma (\tilde c)}
    \int_{p_T^{\rm cut}}^\infty
    \frac{{\rm d} \sigma (\tilde c)}{{\rm d} p_T(\ell^+, \ell^-)} \ 
    {\rm d} p_T(\ell^+, \ell^-) \ .
\label{eff_def} \ee
As our simulation is performed at the leading order accuracy, uncertainties are
expected to be large. Although the $\varepsilon$ quantity exhibits a ratio,
the cancellation of the uncertainties is only partial as the phase space cuts
are different for the numerator and the denominator. More accurate estimates
require the computation of higher-order corrections as well as the resummation
of the Sudakov logarithms that are potentially significant for large $p_T$
values.

On the lower panel of Figure~\ref{fig:ZH}, we investigate the angular
separation of the two leptons and observe that the EFT effects distort the
shape of the spectrum that is more uniform in the Standard Model than when
EFT effects are included. A shape analysis going beyond the scope of this paper,
we instead define the asymmetry
\be
  {\cal A}_{\Delta\tilde{\phi}}(\tilde c) = \frac{
   {\rm d}\sigma\big(\Delta\tilde{\phi}(\ell^{+}, \ell^{-})<0\big) -
   {\rm d}\sigma\big(\Delta\tilde{\phi}(\ell^{+}, \ell^{-})>0\big)}{
   {\rm d}\sigma\big(\Delta\tilde{\phi}(\ell^{+}, \ell^{-})<0\big) +
   {\rm d}\sigma\big(\Delta\tilde{\phi}(\ell^{+}, \ell^{-})>0\big)} \ ,
\label{asym_def}\ee
that we use as a second handle on CPV new physics effects, in addition to the
$\varepsilon$ variable defined by Eq.~\eqref{eff_def}. In the right-hand side of
the above expression, the dependence on the Wilson coefficient is understood for
clarity.

\begin{figure}
  \centering
  \includegraphics[scale=0.72]{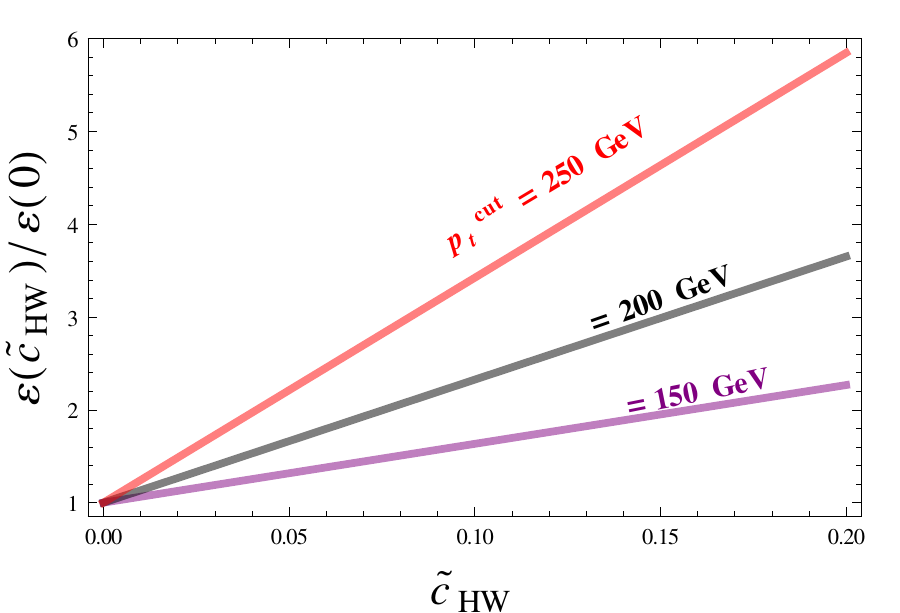}
  \includegraphics[scale=0.191]{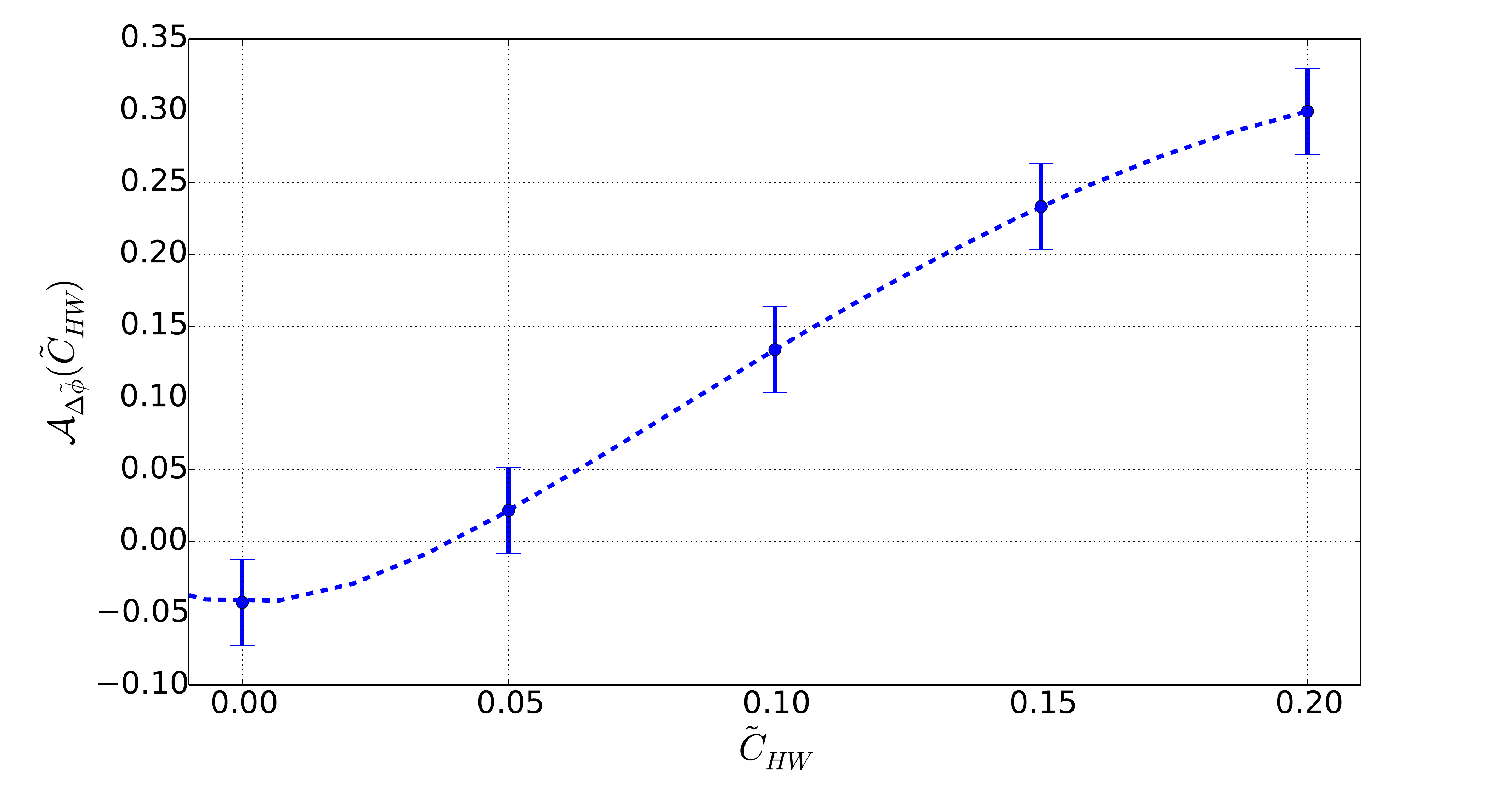}
  \caption{$\chw$ dependence of the $\varepsilon$ variable defined in
   Eq.~\eqref{eff_def} (left) for different choices of the $p_T^{\rm cut}$
   threshold, and of the asymmetry defined in Eq.~\eqref{asym_def} (right).}
  \label{fig:effZH}
\end{figure}

\begin{figure}
  \centering
  \includegraphics[width=0.95\columnwidth]{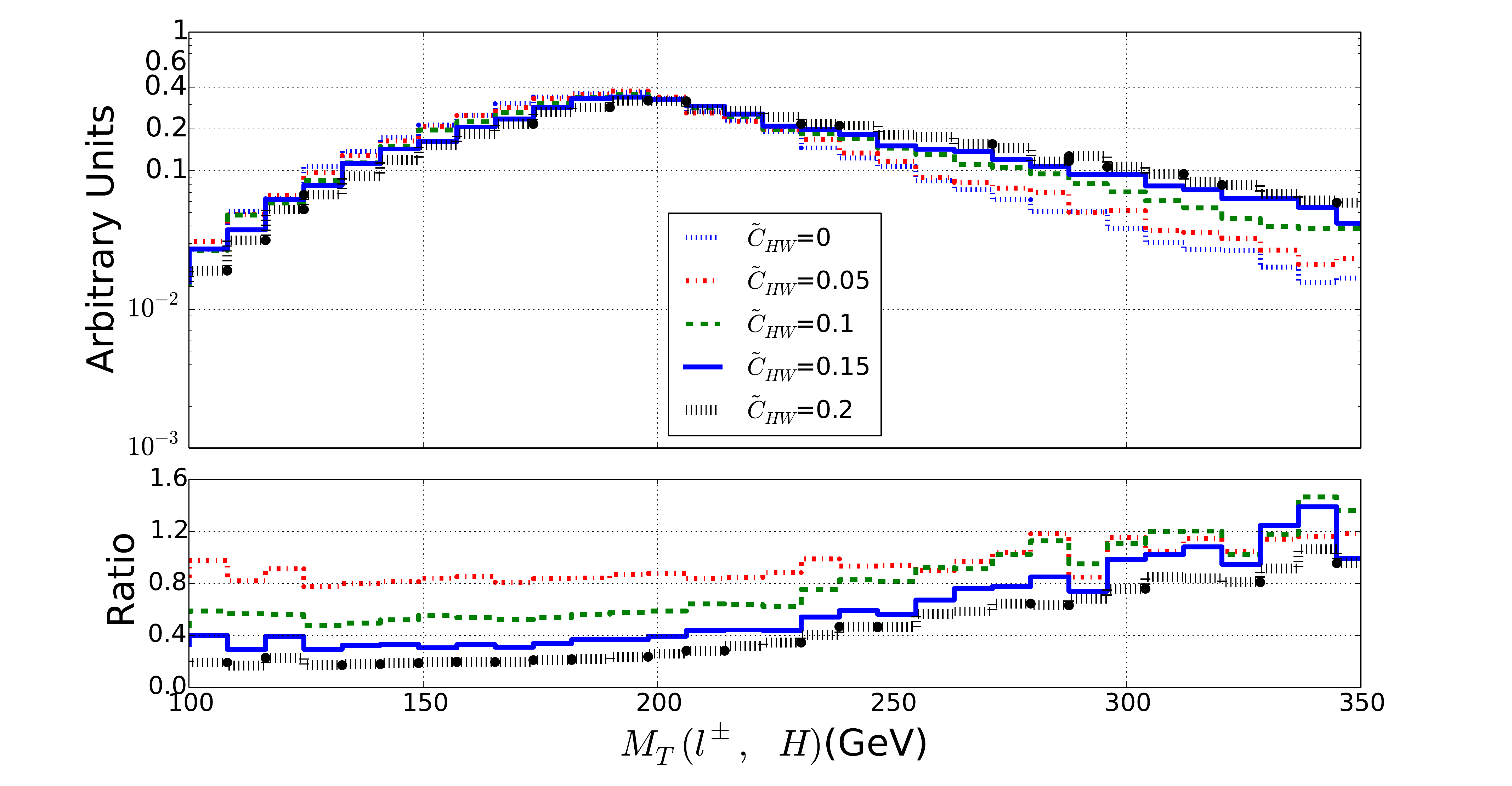}
  \includegraphics[width=0.95\columnwidth]{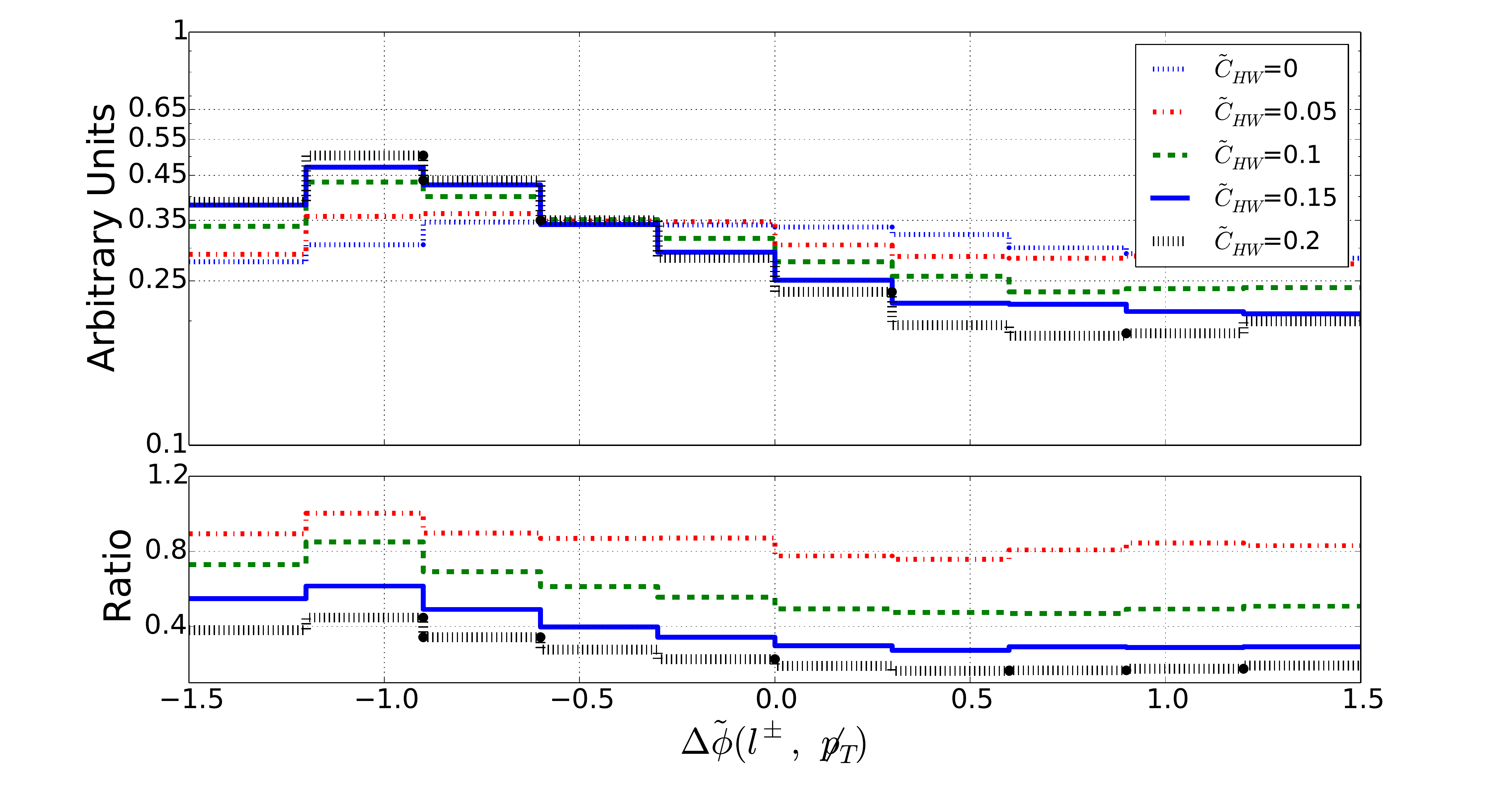}
  \caption{Representative kinematical properties of the decay product of a $WH$
    system produced in LHC collisions at a center-of-mass energy of 13~TeV. We
    consider the transverse mass of the $WH$ system (top) and the
    angular separation in azimuth between the lepton and the missing momentum
    (bottom). We allow for different values for the $\chw$ parameter and we
    present, in the lower panels, the bin-by-bin ratio of the new physics
    predictions to the Standard Model expectation.}
  \label{WH}
\end{figure}

The dependence of the $\varepsilon$ and ${\cal A}_{\Delta\tilde{\phi}}$
observables on the $\chw$ parameters is presented in Figure~\ref{fig:effZH}. As
expected, a harder selection on $p_T(\ell^+, \ell^-)$ implies a larger
sensitivity to the EFT operators through the $\varepsilon$ variable, so that it
offers a way to probe smaller values of the $\chw$ parameter. Conclusive
statements should however also account for the reduction of the fiducial cross
section, and hence depend on the considered luminosity and the appropriately
designed event selection strategy. The
${\cal A}_{\Delta\tilde{\phi}}$ asymmetry moreover shows that large deviations
from the Standard Model could be expected, including a possible different sign
for some $\chw$ values.
Measuring  such an observable with a reasonable precision could therefore yield
an extra way to constrain EFT deviations.

The Higgs boson could also be produced in association with a $W$-boson, which
leads to a final state containing a single lepton once a $W$-boson leptonic
decay is accounted for. We again construct appropriate observables that
allow for the extraction of bounds on the EFT parameters. In Figure~\ref{WH},
we show, in the upper panel, the distribution in the transverse mass of the
lepton and the reconstructed Higgs boson system, $M_T(\ell, H)$, and the angular
separation in azimuth between the lepton and the missing transverse momentum
$\Delta\tilde\phi(\ell, \slashed{p}_T)$ (lower panel), this last observable
being defined similarly to Eq.~\eqref{dphitilde}.

\begin{figure}
  \centering
  \includegraphics[scale = 0.72]{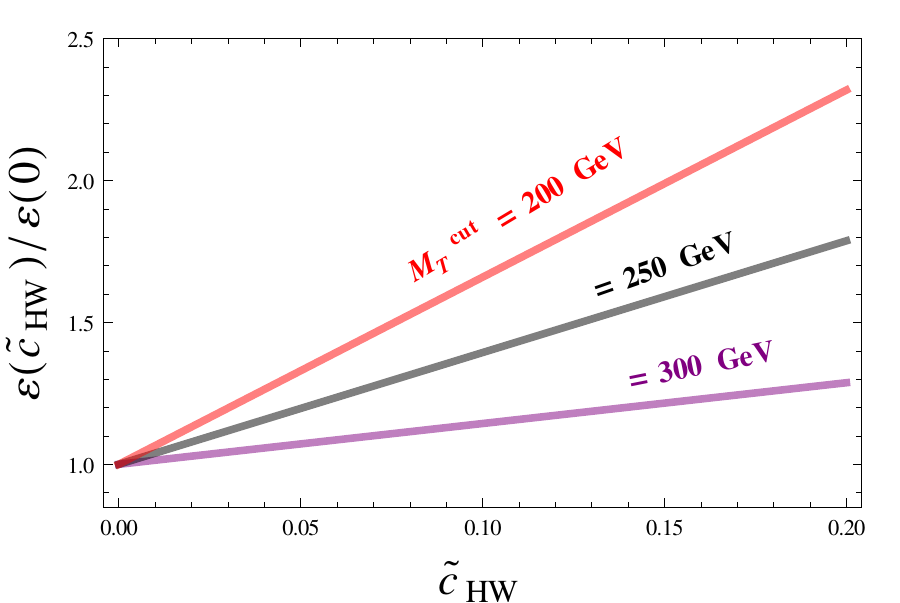}
  \includegraphics[scale = 0.191]{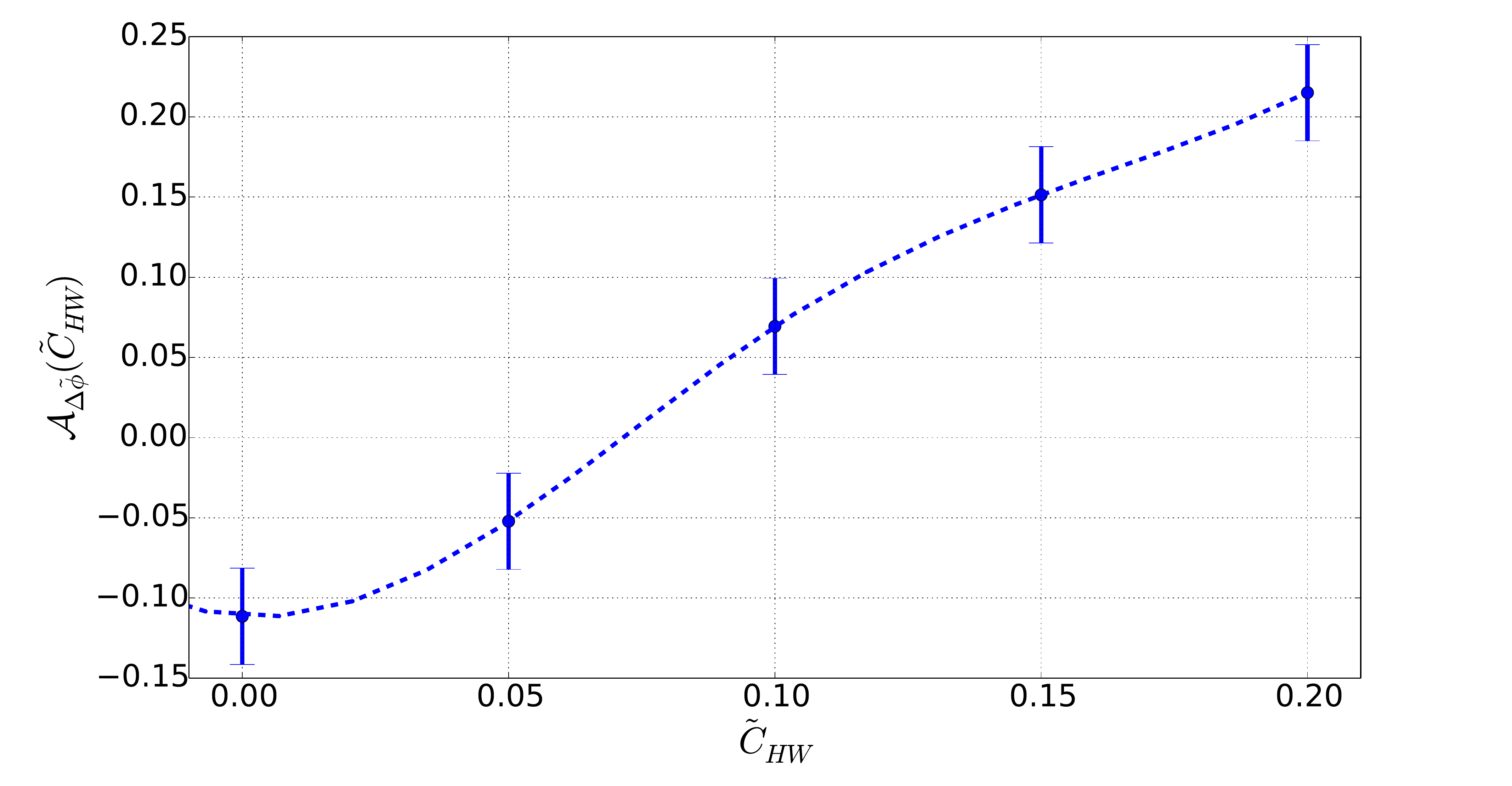}
  \caption{Same as in Figure~\ref{fig:effZH} but for $WH$ production.}
  \label{fig:effWH}
\end{figure}

We observe effects that are similar to the $ZH$ case, the EFT operators under
consideration impacting the tail of the invariant mass distribution whose
fall at large $M_T(\ell, H)$ values is tamed and yielding a more pronounced
shape for the $\Delta\tilde\phi(\ell, \slashed{p}_T)$ spectrum. We define an
$\varepsilon$ efficiency analogously to Eq.~\eqref{eff_def},
\be
  \varepsilon({\tilde c}) = \frac{1}{\sigma (\tilde c)}
    \int_{M_T^{\rm cut}}^\infty
    \frac{{\rm d} \sigma (\tilde c)}{{\rm d} M_T(\ell, H)} \ 
    {\rm d} M_T(\ell, H) \ ,
\ee
which now depends on the Wilson coefficients and on the $M_T^{\rm cut}$ minimum
value for the transverse mass, as well as an asymmetry as in
Eq.~\eqref{asym_def},
\be
  {\cal A}_{\Delta\tilde{\phi}}(\tilde c) = \frac{
   {\rm d}\sigma\big(\Delta\tilde{\phi}(\ell^\pm, \slashed{p}_T)<0\big) -
   {\rm d}\sigma\big(\Delta\tilde{\phi}(\ell^\pm, \slashed{p}_T)>0\big)}{
   {\rm d}\sigma\big(\Delta\tilde{\phi}(\ell^\pm, \slashed{p}_T)<0\big) +
   {\rm d}\sigma\big(\Delta\tilde{\phi}(\ell^\pm, \slashed{p}_T)>0\big)} \ .
\ee
we obtain the results represented in Figure~\ref{fig:effWH} from which we
observe that All $VH$ modes offer
extra means to constrain CPV operators, the $WH$ channel
however benefiting from a larger cross section so that it could be in principle
more promising.

\subsection{Higgs production by vector boson fusion}
\label{sec:VBF}

\begin{figure}
  \centering
  \includegraphics[width=0.75\columnwidth]{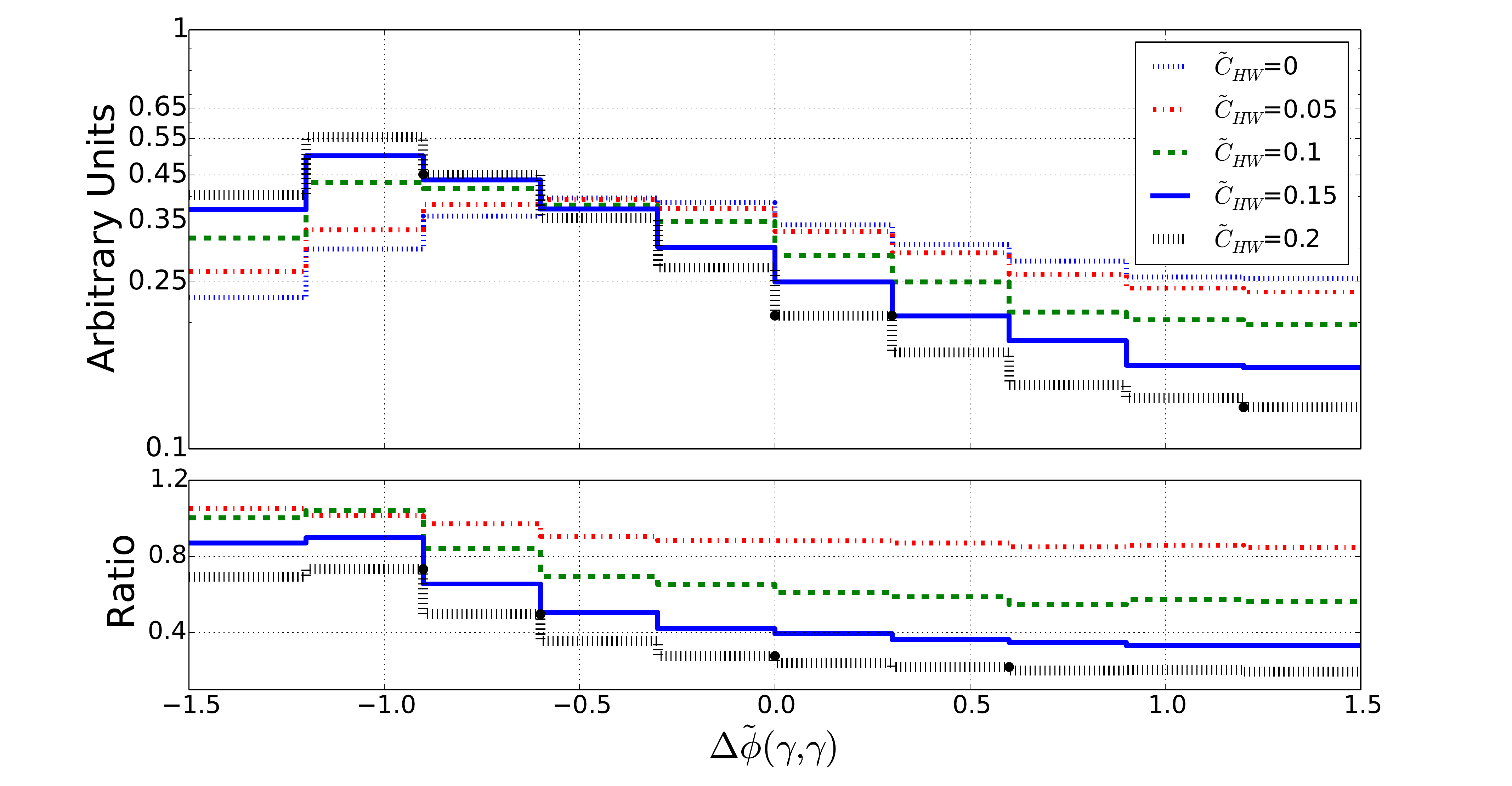}\\
  \includegraphics[width=0.75\columnwidth]{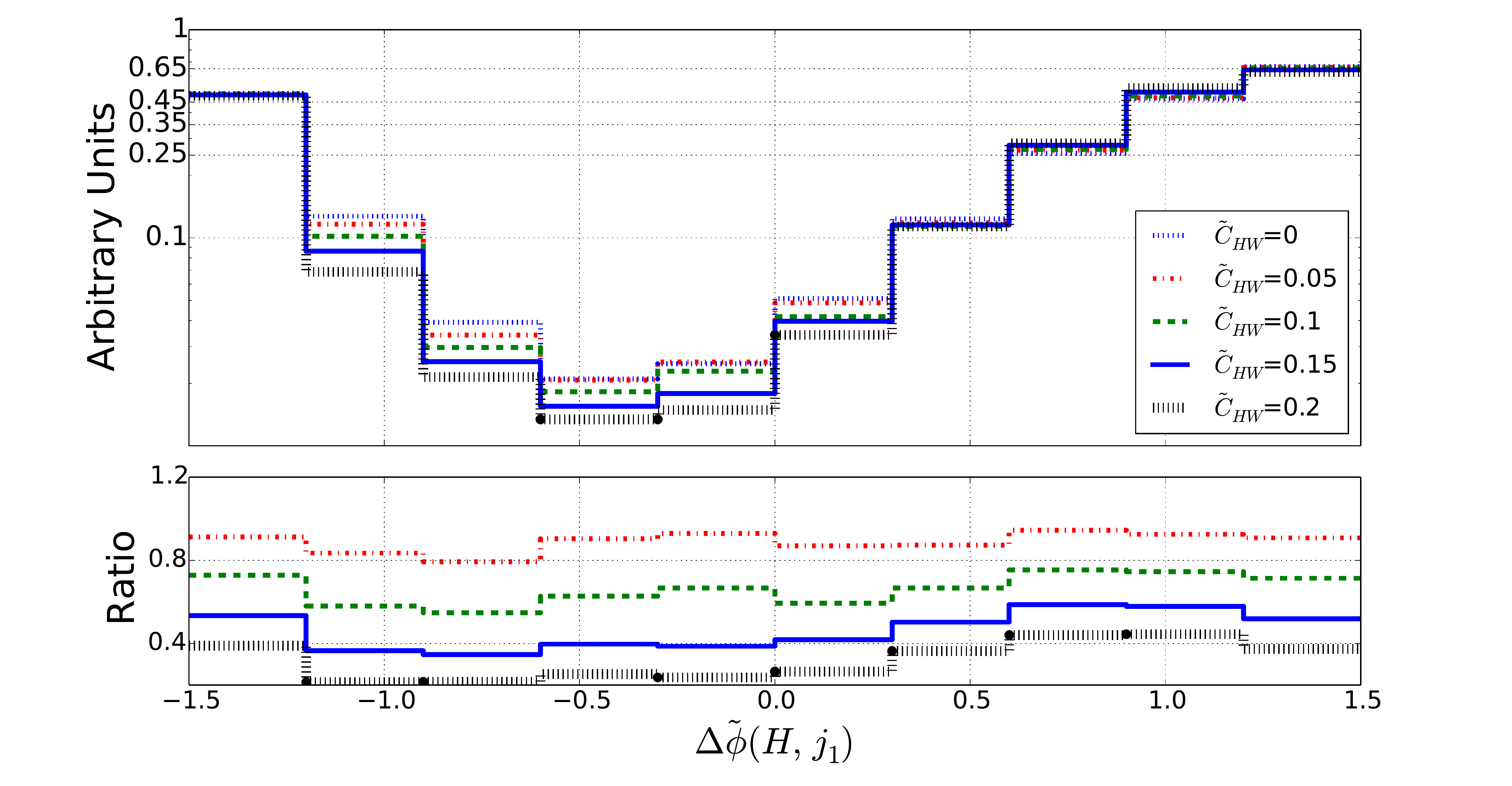}\\
  \includegraphics[width=0.75\columnwidth]{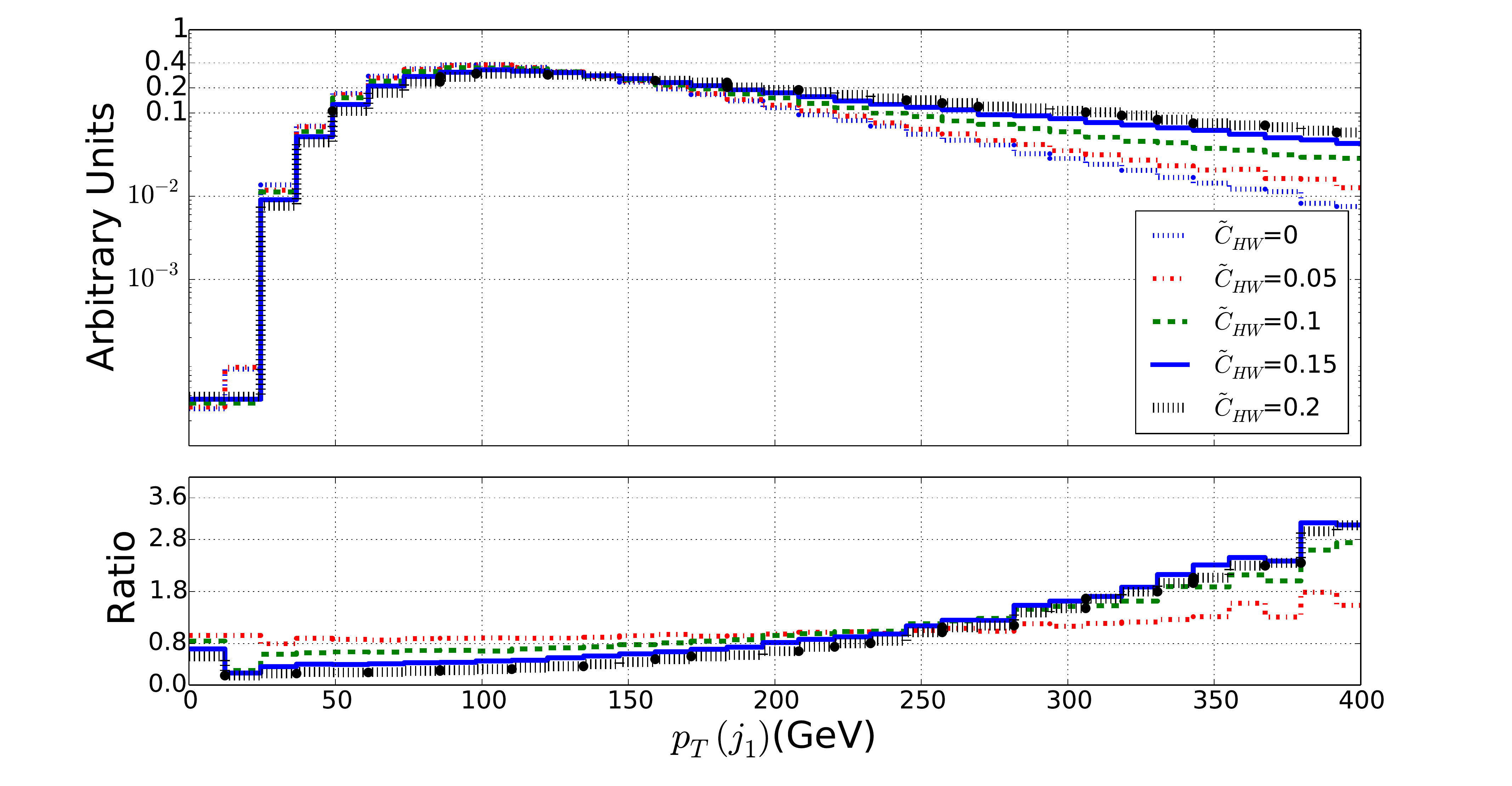}
  \caption{Representative kinematical properties of the decay product of a Higgs
    boson produced by vector boson fusion and that decays into a photon pair
    when produced in LHC collisions at a center-of-mass energy of 13~TeV. We
    consider the angular separation in azimuth between the photon pair
    originating from the Higgs boson decay (top), the angular separation in
    azimuth between the reconstructed Higgs boson and the leading jet (center)
    and the transverse momentum of the leading jet (bottom). We allow for
    different values for the $\chw$ parameter and we present, in the lower
    panels, the bin-by-bin ratio of the new physics predictions to the Standard
    Model expectation.}
  \label{fig:vbf}
\end{figure}

Vector boson Higgs boson production processes are excellent probes of
physics beyond the Standard Model, in particular when new physics is
parameterized within the EFT framework. We focus on three variables which we
have found very sensitive to CPV EFT operators, namely the angular separation
in the transverse plane $\Delta\tilde\phi(\gamma, \gamma)$, between the decay
products of the Higgs boson (considered to be a photon pair), the
transverse momentum of the leading forward jet $p_T(j_1)$ and the angular
separation in the transverse plane $\Delta\tilde\phi(H, j_1)$ between the
reconstructed Higgs boson and the leading forward jet.
The distributions in these three observables are shown in Figure~\ref{fig:vbf},
where we observe a standard EFT behaviour. The transverse-momentum spectrum of
the leading forward jet departs from the Standard Model expectation for large
$p_T$
values, the distribution being then harder, and the shapes of the two angular
variable distributions is distorted, the effects being more pronounced for
$\Delta\tilde\phi(\gamma, \gamma)$. We have verified that these effects are also
observed in observables for which we have not presented the results, like the
distribution in the transverse momentum of the Higgs boson $p_T (H)$ that is
actually strongly correlated to the one of the leading forward jet. The
enhancement in the tail of the spectrum is moreover also correlated with the
suppression of events featuring a large angular separation. Additional
information can be obtained by studying the $\Delta \tilde \phi(H, j_1)$
spectrum for $\Delta \tilde \phi$ values in the [-1.25, 0.25] range.

\begin{figure}
  \centering
  \includegraphics[scale = 0.72]{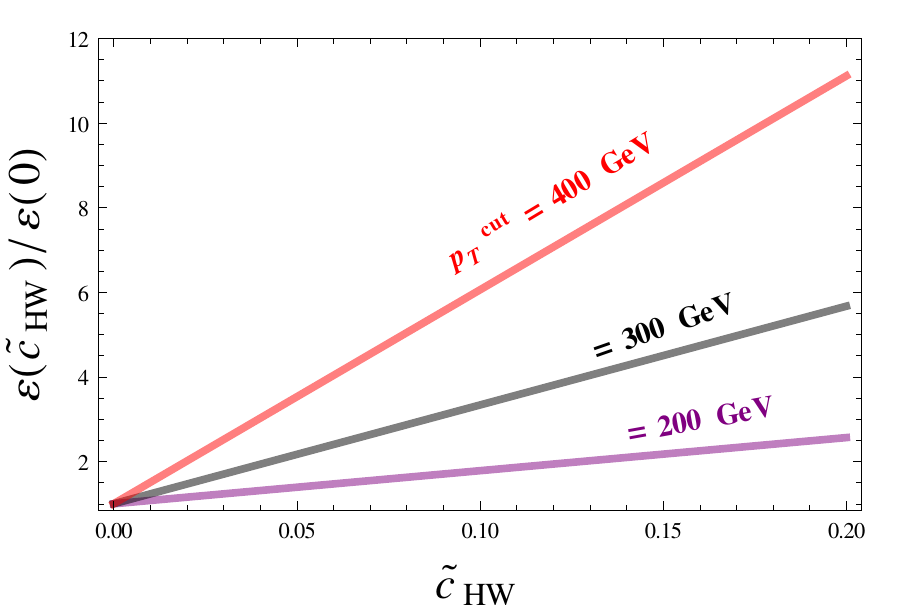}
  \includegraphics[scale = 0.191]{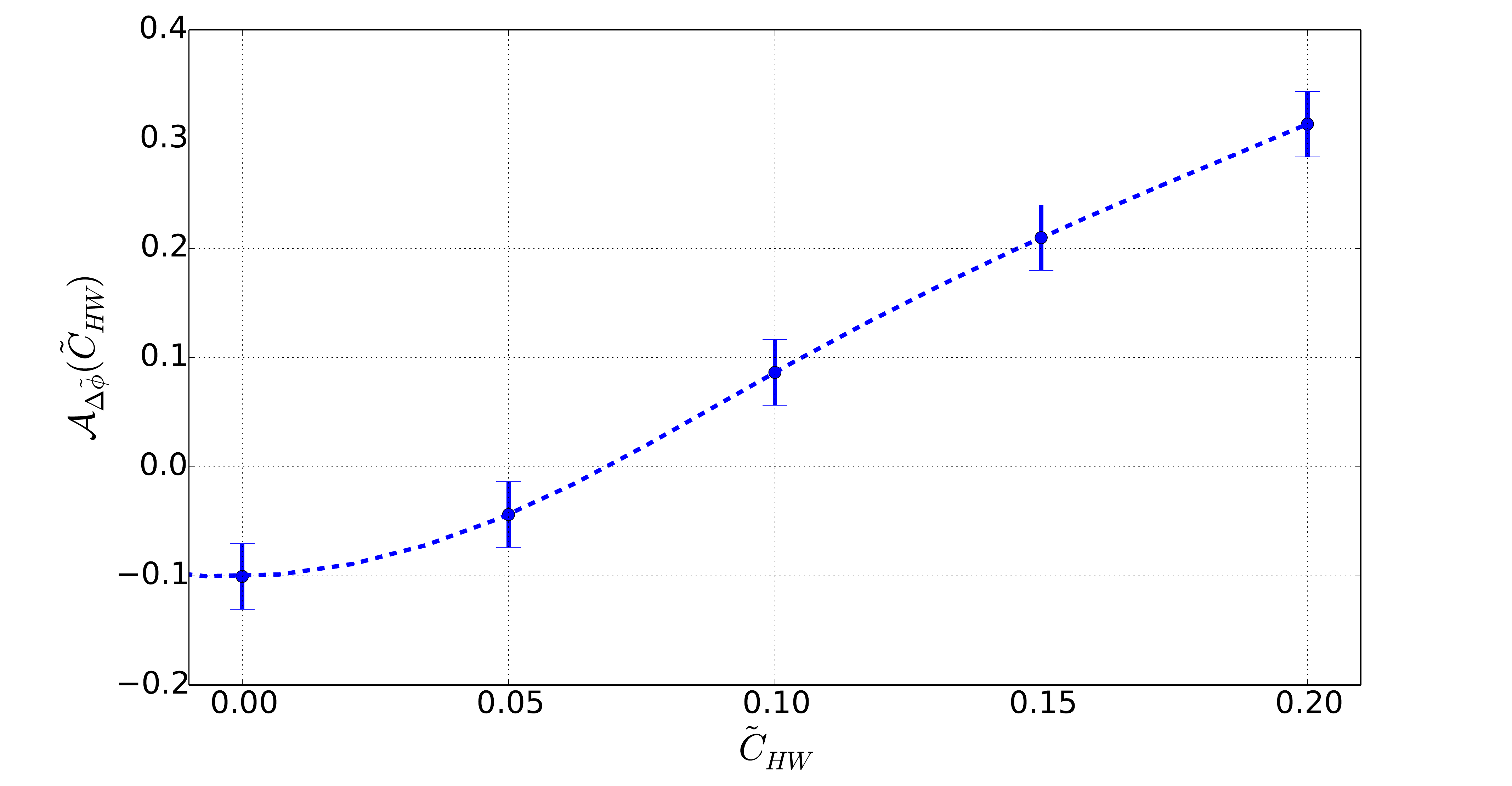}
  \caption{Same as in Figure~\ref{fig:effZH} but for VBF Higgs-boson production
    and the observables considered in Section~\ref{sec:VBF}.}
  \label{eff:VBF}
\end{figure}

We define asymmetries (for the angular variables) and efficiencies (for the
dimensionful variable) as in the previous section so that these observable can be
used for extracting constraints on EFT operators. This is confirmed by the
results presented in Figure~\ref{eff:VBF}. We have in particular found a
stronger dependence of the asymmetry connected to the Higgs-boson decay
products.

\subsection{CPV EFT effects in dileptonic $W$-boson pair production events}

While all previously considered processes allow us to get information on the
${\cal O}_g$, ${\cal O}_\gamma$, ${\cal O}_{HW}$ and ${\cal O}_{HB}$ operators,
the ${\cal O}_{3W}$ operator can instead only be constrained by the study of
$W$-boson pair production, as already shown in Section~\ref{sec:Run1} and
Section~\ref{sec:future}. We focus on a final state signature made of two
leptons and missing energy, each $W$-boson hence decaying leptonically.
After examining several distributions, we have found that the EFT effects are
particularly important in the distribution in the invariant mass of the dilepton
system $M(\ell^+ \ell^-)$, as well as in an analogous of the $\mathcal{O}_{1}$
observable introduced in the context of four-leptonic decays of the Higgs
boson~\cite{Han:2009ra,Christensen:2010pf},
\be
  \tilde{\mathcal{O}}_{1} = \frac{\mathbf{p}_{+}\times \mathbf{p}_{-}}
    {\vert\mathbf{p}_{+}\times \mathbf{p}_{-}\vert}\
    {\rm sign}\big[(\mathbf{p}_{+}-\mathbf{p}_{-}) \cdot \hat{\mathbf z}\big]\ ,
\label{eq:o1}\ee
where $\mathbf{p}_{\pm}$ denotes the three-momentum of the lepton $\ell^\pm$ and
$\hat{\mathbf z}$ is a unit vector along the collision axis.

\begin{figure}
  \centering
  \includegraphics[width=0.95\columnwidth]{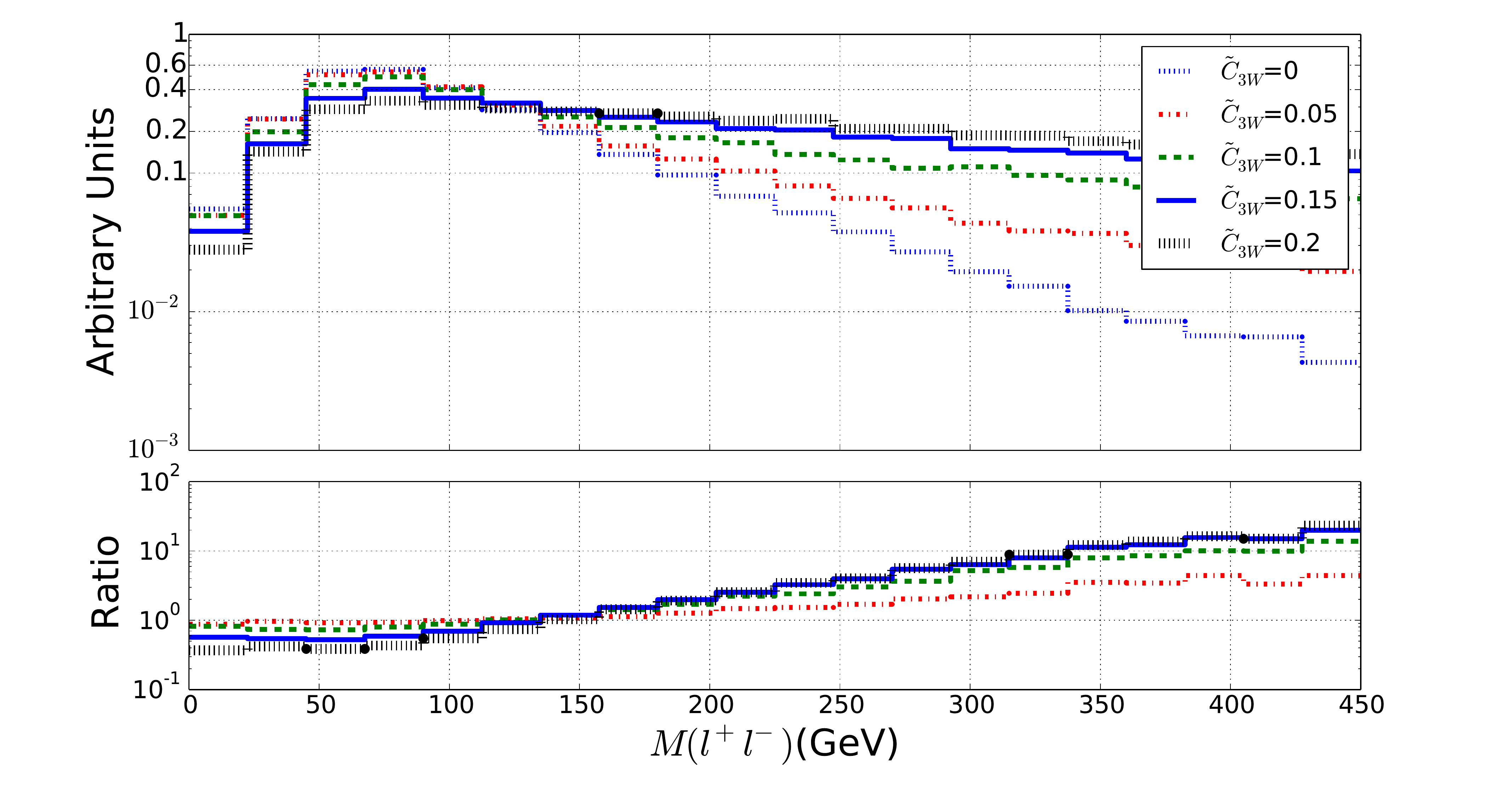}
  \includegraphics[width=0.95\columnwidth]{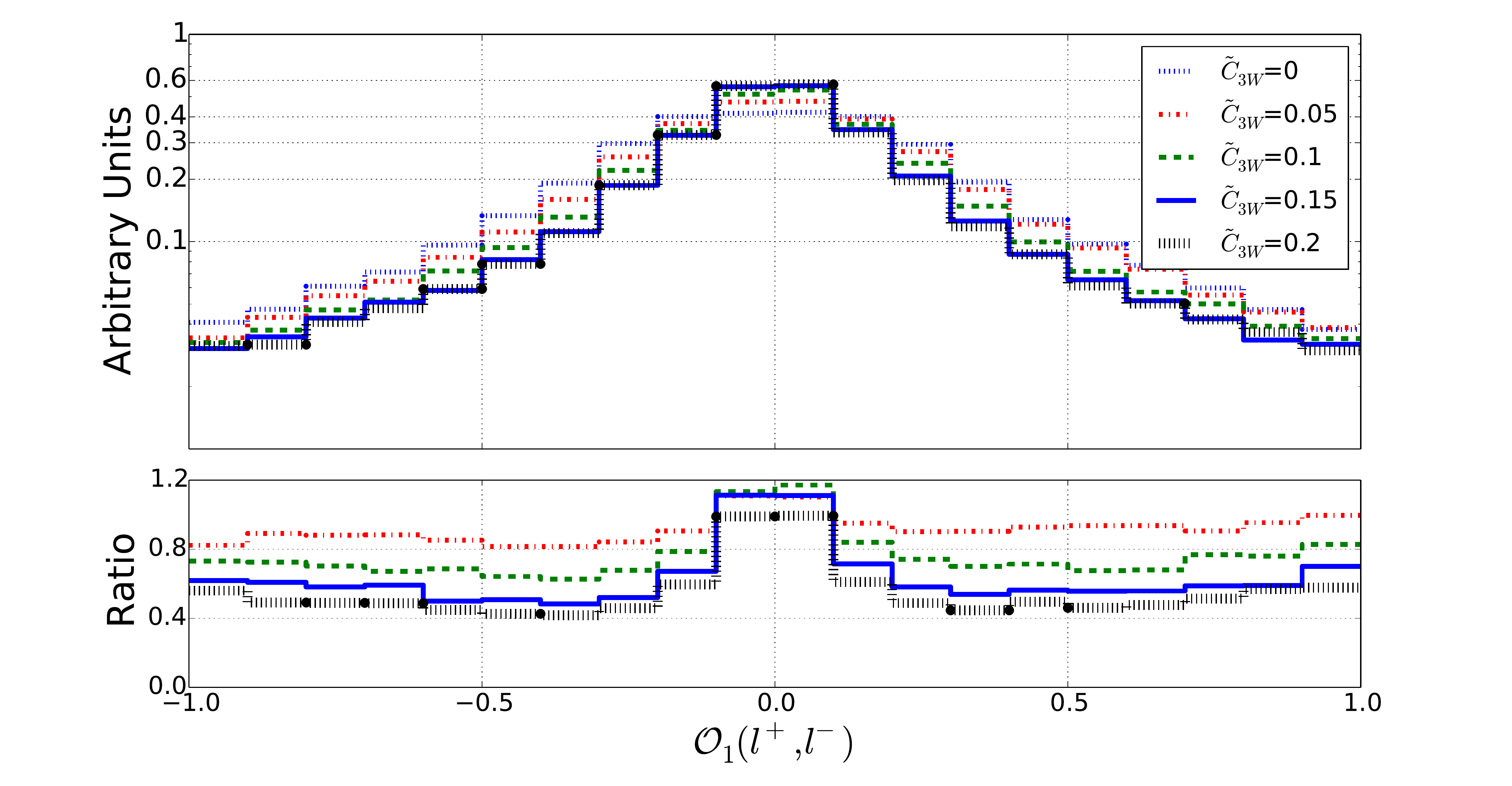}
  \caption{Representative kinematical properties of the decay products of a
    $W$-boson pair produced in LHC collisions at a center-of-mass energy of
    13~TeV. We consider the invariant mass of the dilepton pair issued from the
    $WW$ system (top) and the $\tilde{\mathcal O}_1$ observable defined by
    Eq.~\eqref{eq:o1} (bottom). We allow for different values for the $\ctw$
    parameter and we present, in the lower panels, the bin-by-bin ratio of the
    new physics predictions to the Standard Model expectation.}
  \label{fig:ww}
\end{figure}

We present predictions for the two selected observables in Figure~\ref{fig:ww}
for different values of the $\ctw$ Wilson coefficient. Once again, the tail of
the spectrum in the dimensionful $M(\ell^+ \ell^-)$ variable turns out to be
very sensitive of EFT effects, the distribution becoming harder, and the shape
of the spectrum in the
$\tilde{\mathcal{O}}_{1}$ observable is modified with respect to the Standard
Model case. Similarly to the previous section, we could encapsulate these
differences in the definition of an efficiency and an asymmetry that would
provide handles on the effective parameters.

\subsection{Revisiting CPV Higgs-boson studies in the four-lepton final state}

Traditionally, studies of $CP$ violation in the Higgs sector have been mostly
focused on the
four-lepton final state originating from a Higgs-boson decay into a $Z$-boson
system~\cite{Choi:2002jk,Godbole:2007cn,Hagiwara:2009wt,Englert:2010ud,%
Chen:2014ona}. In this
section, we revisit those studies and show how including appropriate selections
could enhance the sensitivity to the EFT operators of the Lagrangian of
Eq.~\eqref{LCPV}.
We start our analysis by performing an event selection that requires the
presence of two pairs of leptons with an opposite electric charge. The invariant
mass of the first lepton pair denoted by $Z_1$ is imposed to lie in the
$[75, 105]$~GeV range, whilst the one of the second lepton pair denoted by $Z_2$
is enforced to be included in the
$[10, 200]$~GeV mass window. The first lepton pair is hence identified with an
on-shell $Z$-boson, and the second pair corresponds to the off-shell $Z$-boson
issued from the Higgs-boson decay.

\begin{figure}
  \centering
  \includegraphics[width=0.95\columnwidth]{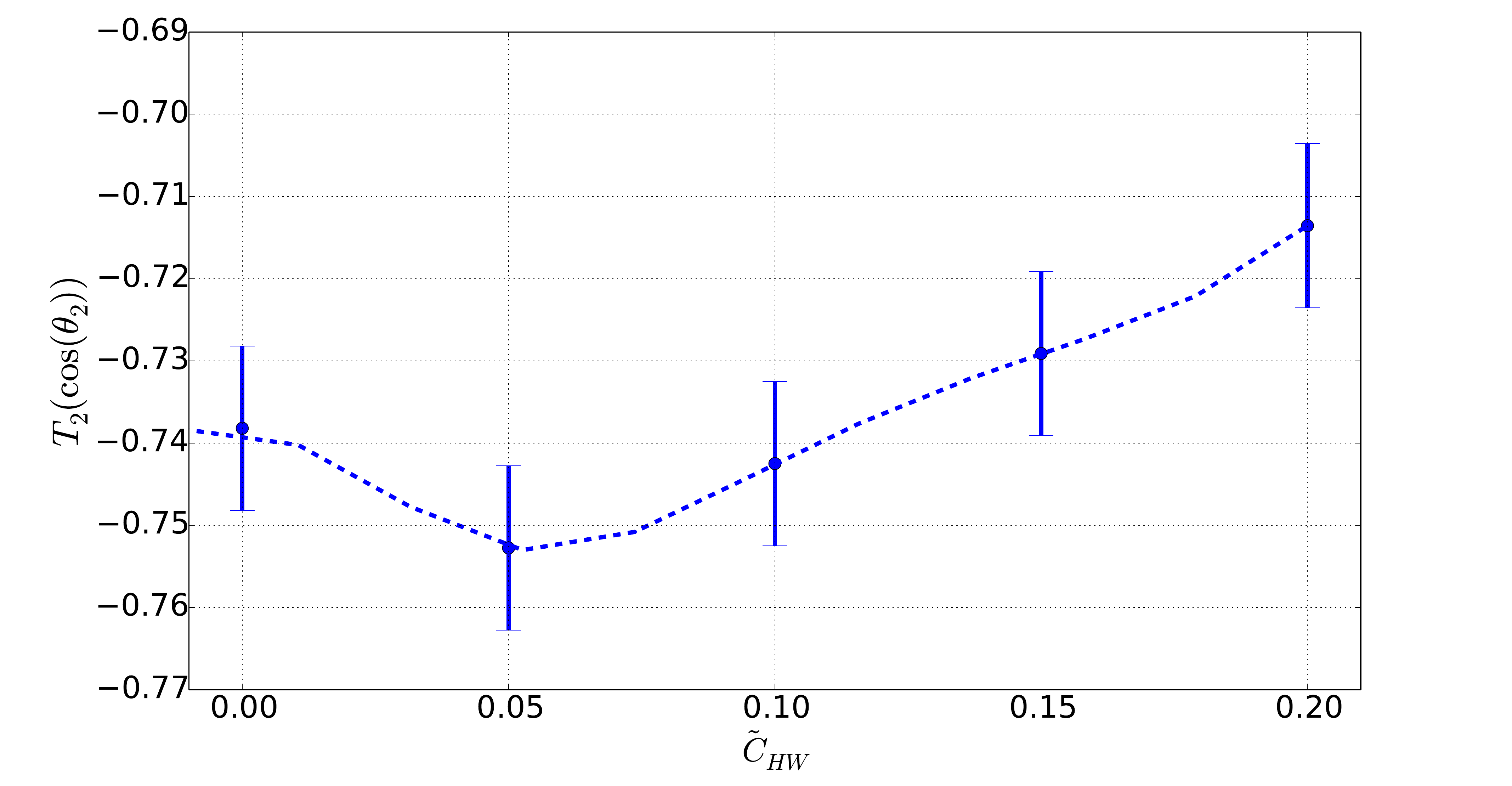}
  \includegraphics[width=0.95\columnwidth]{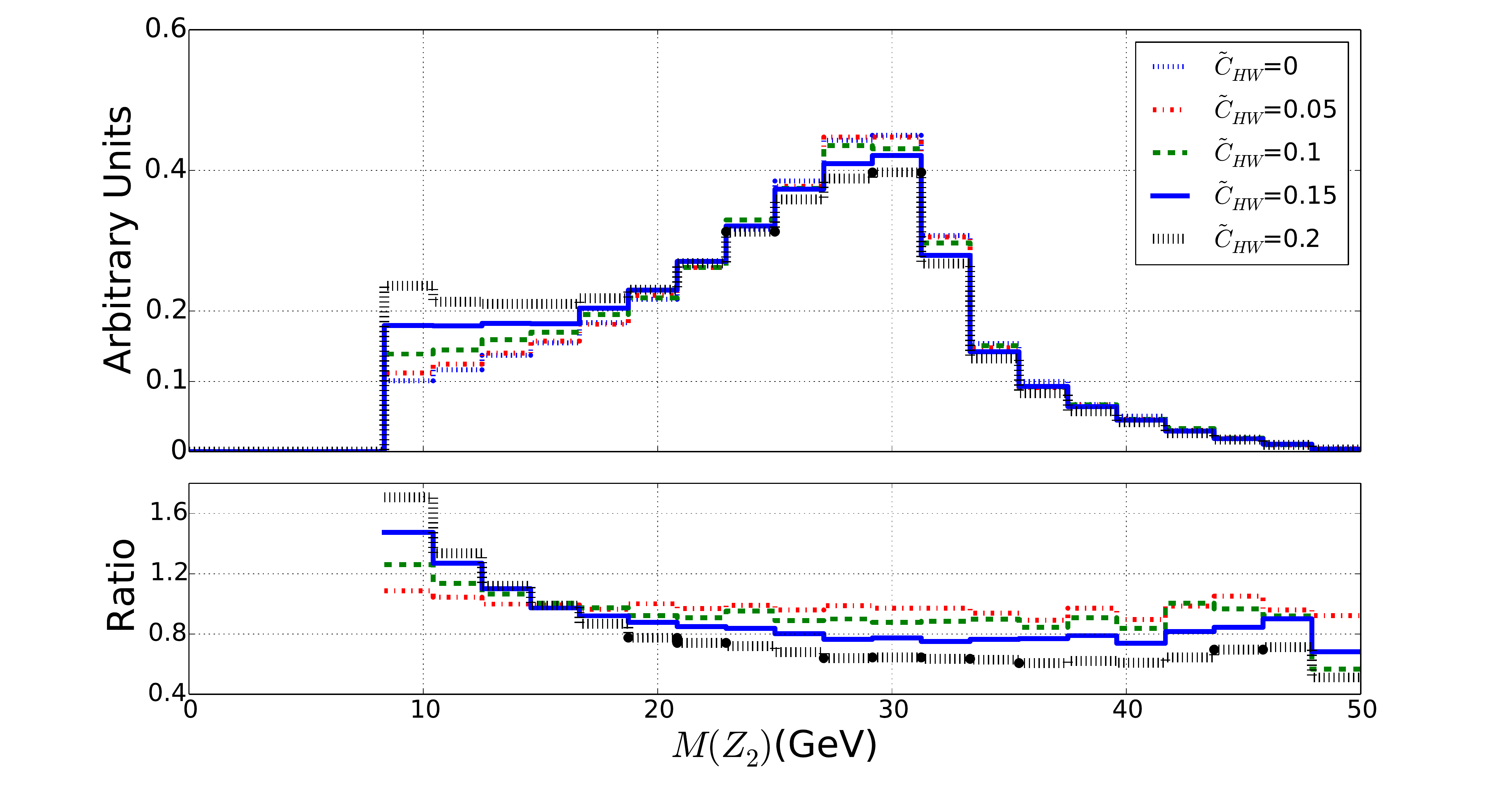}
  \caption{Representative kinematical properties of the four lepton system
    originating from a Higgs boson that is decaying into a $Z$-boson pair and
    that has been produced in LHC collisions at a center-of-mass energy of
    13~TeV. We
    consider the $T_2(\cos\theta_2)$ variable as defined in the text and present
    its dependence on the $\chw$ parameter, together with the one of the
    off-shell $Z$-boson
    invariant mass distribution (bottom), for varied $\chw$ values. In this last
    case, we also show, in the lower inset of the figure, the bin-by-bin ratio
    of the new physics predictions to the Standard Model expectation.}
    \label{fig:4L}
\end{figure}

Key observables for CPV studies include the polar angles of the leptons,
$\theta_{1}$ and $\theta_{2}$, evaluated in the rest frame of the parent $Z_1$
and $Z_2$ bosons, as well as the azimuthal angle $\varphi$ between the two
planes formed by the lepton pairs in the Higgs-boson rest frame. Exploring the
traditional variables, we have observed that a particular function of the lepton
polar angles,
\be\bsp
  &\ T_2(x) = \frac43 \Big[{\rm d}\sigma(-1<x<-1/2)\\ &\ \
    -{\rm d}\sigma(-1/2<x<1/2)
    +{\rm d}\sigma (1/2<x<1)\Big] \ ,
\esp\ee
(with $x=\cos\theta_1$ or $\cos\theta_2$) is very sensitive to the presence of
EFT operators. The $T_2(\cos\theta_2)$ dependence on $\chw$ is presented in
Figure~\ref{fig:4L} (upper panel) for illustrative purposes. In this example, we
observe a $\chw$ dependence that could possibly be exploited by precise
measurements. We additionally show, in the lower panel of the figure, the
invariant-mass distribution of the $Z_2$ system that additionally feature a
dependence on the EFT parameters and could provide an extra handle to better
corner deviations from the Standard Model.

\section{Discussion}\label{sec:discussion}

We have attempted to find new avenues for probing the impact
of possible $CP$-odd interactions of the Higgs boson. We have considered two
different approaches. First, we have made use of total rate measurements to both
evaluate the current status of the constraints on all bosonic effective $CP$-odd
operators and their prospects. Second, we have considered pairs of observables
that allows in principle to get a joint sensitivity to the EFT and CPV effects.
One observable is dimensionful so that large momentum transfers could be probed,
and another observable involves angles so that the CPV impact is expected to be
significant.

We have shown that the constraints that can be derived on the basis of the Run~I
LHC cross section results will only be barely improved during the next 20 years.
Going beyond the total rate approach is thus mandatory in order to corner the
Higgs sector better. Differential distributions are powerful handles
for a variety of processes. We recast the dimensionful observable as an
efficiency of selecting a part of the phase space where the observable under
consideration satisfies some condition. On the other hand, the angular
observable is connected to an asymmetry.

Our findings can be summarized as
follows.
\begin{itemize}
\item {\bf{VH production}}: The dimensionful observable is ta\-ken to be the
  scalar sum of the transverse momenta of the two leptons originating from the
  decay of $Z$-boson in the $ZH$ case, and the transverse mass of the system
  comprised of the reconstructed Higgs boson and the lepton issued from the
  $W$-boson decay in the $WH$ case. The angular observable is taken to be the
  difference in azimuthal angle between the two leptons (the lepton and the
  missing
  momentum) in the $ZH$ ($WH$) case. We have found that this efficiency and the
  asymmertry built from the angular observable provide an effective handle to
  distinguish CPV effects 
\item {\bf{VBF production}}: Similarly, we make use of the azimuthal angular
  separation of the diphoton system arising from the Higgs boson decay and the
  transverse momentum of the leading jet.
\item  {\bf{Dileptonic $W$-boson pair production}}: Here, we use the invariant
  mass of the dilepton system for computing the efficiency related to the
  dimensionful observable, and the triple product observable
  $\tilde{\cal{O}}_{1}$ as a dimensionless variable.
\item {\bf {Higgs decays in four lepton final state}}: In this case, we rely on
  the reconstructed off-shell $Z$-boson stemming from the Higgs boson decay. We
  consider its invariant mass as a dimensionful variable, and the so-called
  $T_2$ function applied on the polar angles of its decay products as the
  dimen sionless variable.
\end{itemize}

In order to be able to compare the sensitivity expected by the usage of
pairs of observables with respect to the use of cross section measurements,
there are two ways. Either we need to rely on the corresponding experimental
studies, that are not performed yet, or we need to perform ourselves the
simulation of both the signal and the
Standard Model background including the parton shower and hadronization effects,
as well as the simulation of the impact of the detector response.

\begin{figure}
  \centering
  \includegraphics[width=0.95\columnwidth]{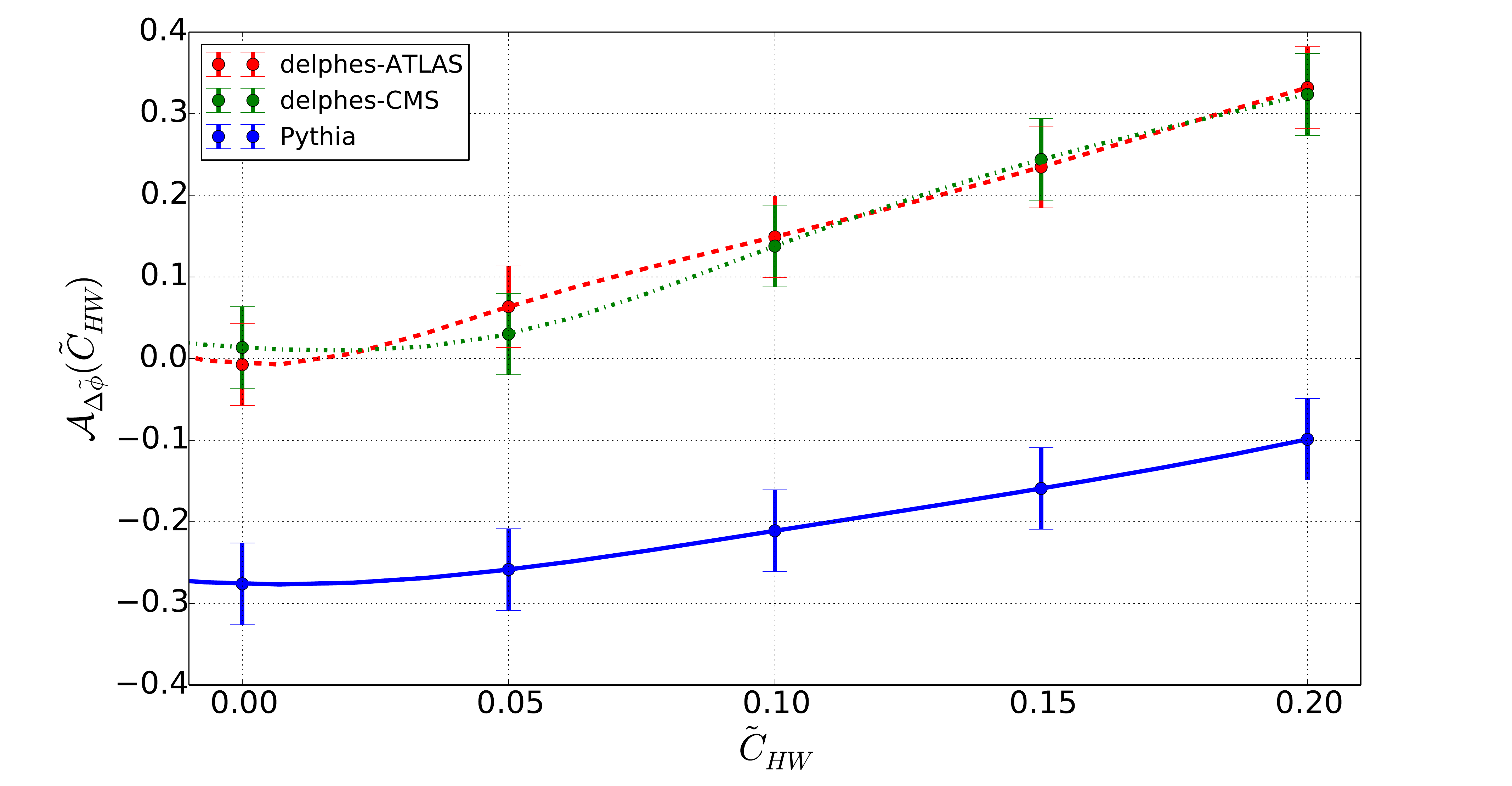}
  \includegraphics[width=0.95\columnwidth]{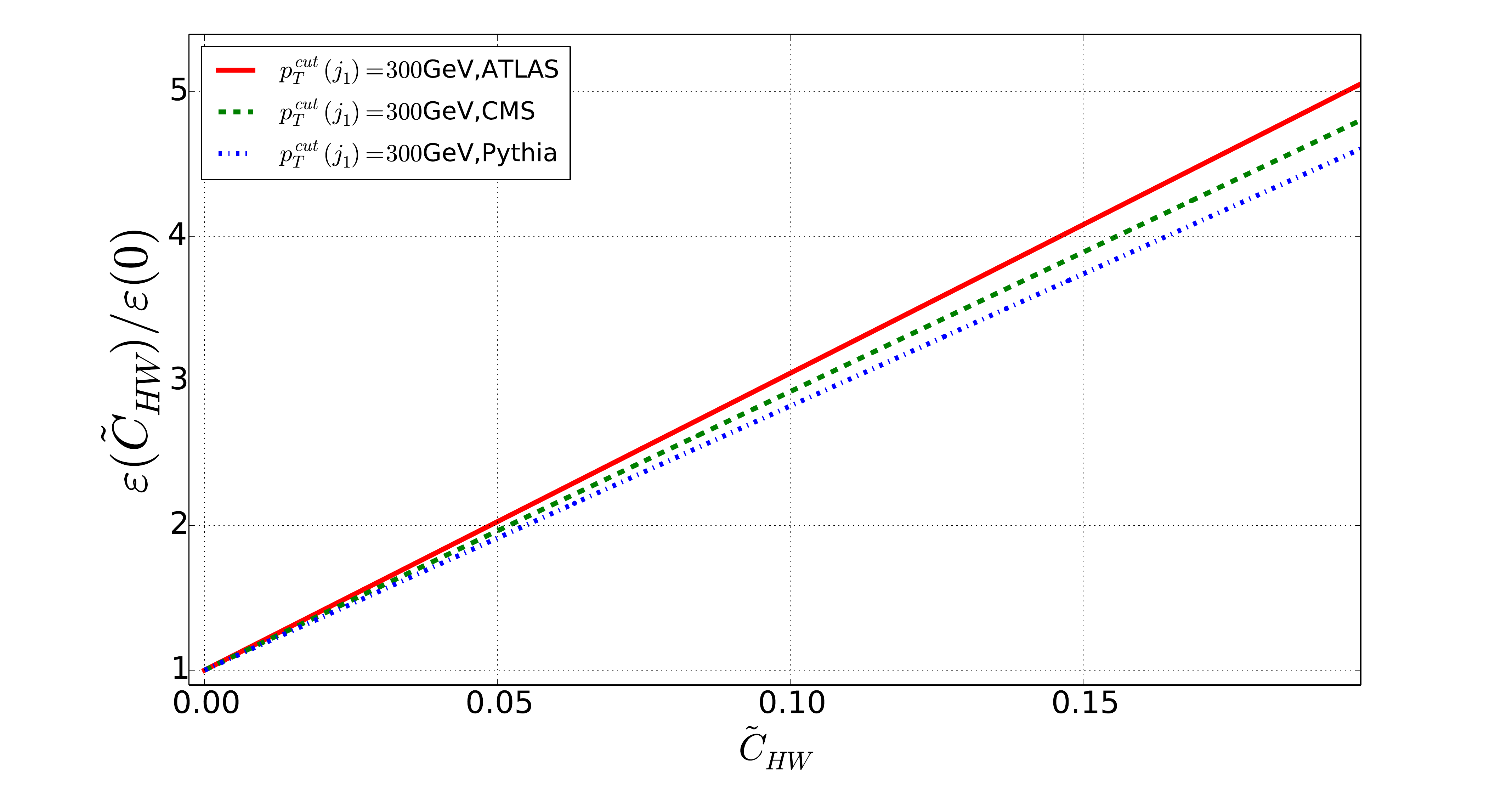}
  \caption{Evaluation of the detector impact on the asymmetry (top) and
    efficiency (bottom) introduced in the context of VBF Higgs boson production
    and defined in Section~\ref{sec:VBF}. We compare predictions solely
    including parton shower and hadronization effects (blue) to predictions
    embedding the modelling of the ATLAS (red) and CMS (green) detector effects.}
  \label{fig:pythia}
\end{figure}

As a first step in the second direction, we evaluate in
Figure~\ref{fig:pythia} the effects that could stem from the parton showering
and hadronization as modelled by {\sc Pythia}~\cite{Sjostrand:2006za}, and
those related from the modelling of the ATLAS and CMS detectors as implemented
in {\sc Delphes}~\cite{deFavereau:2013fsa}. In all cases, object
reconstruction is performed by using the anti-$k_T$ jet
algorithm~\cite{Cacciari:2008gp} as implemented in
{\sc FastJet}~\cite{Cacciari:2011ma}. We present results for the two observables
introduced in the context of VBF Higgs-boson production in
Sec\-tion~\ref{sec:VBF}. Whereas the
$\varepsilon$ efficiency is barely sensitive to detector effects that impact the
results by only a few percents, drastic changes are induced in the distribution
of the ${\cal A}_{\Delta\tilde\phi}$ observable. Additionally, we also observe
significant changes in the normalization with respect to the parton-level
results of Sec\-tion~\ref{sec:VBF}, but the shape dependence on the Wilson
coefficient remains unaltered. It turns out to be even
more pronounced when the detector simulation is included, which reinforces the
motivation for using this variable to characterize new physics in an EFT
context.

The above observables thus require a dedicated study to be performed with the armory of full experimental 
set up including dedicated high $p_{T}$ triggers and a full data driven
background analysis.

\section{Conclusion}
\label{sec:conclusions}

In this paper, we have investigated novel ideas to look for CPV new physics
effects arising both in the couplings of the Higgs boson to the weak vector
bosons and in the self-interactions of the latter. In order to assess those
effects, we have performed an analysis in the context of an effective field
theory once the higher-dimensional part of the Lagrangian is restricted to
relevant CPV operators. We have studied the impact of these new physics EFT
operators on both total rates and differential distributions, as the effects are
known to be larger for processes involving large momentum transfer.

We have first used LHC Run~I data to define the range in which the considered
Wilson coefficients are allowed to vary on the basis of total rate information.
We have then explored the prospects for the next runs of the LHC when we
restrict the analysis to the usage of similar techniques. The expected
improvements have been found rather mild, so that we have investigated how the
use of differential information could play a more important role for maximizing
the potential of future LHC data.

We have more precisely examined a variety of Higgs and electroweak boson
production channels to evaluate the sensitivity of the LHC to new CPV effective
operators. Our analysis has included a focus on the associated production of a
Higgs and a weak boson ($VH$), Higgs-boson production by
vector-boson fusion (VBF), $W$-boson pair production ($W^+W^-$) and the
four-lepton channel traditionally used for CPV Higgs-boson studies. In each
case, we have studied various kinematic distributions and we have selected
the most sensitive ones to EFT effects. We have further proposed several
dimensionless (angular) and dimensionful observable that could be used,
possibly jointly, as novel handles to pin down new physics.

In this work, we have undertaken, as a pioneering study of these new
observables, a beyond the Standard Model signal analysis at the leading-order
accuracy in QCD after matching the fixed-order results to parton showers. A more
precise assessment on the LHC sensitivity to CPV EFT operators through the use
of the new variables that we have proposed however necessitates, on the one
hand, a full signal and Standard Model background analysis for different
luminosity goals and after including the simulation of detector effects. On the
other hand, it is also mandatory to evaluate the impact of higher-order
corrections to the signal. 

The analysis of the background effects and the design of a signal and background
analysis is left for future work, assuming that the signal considered in this
work are sufficiently distinguishable from the Standard Model (as
it has so far been the case). Other aspects could also be investigated in the
future, like the determination (and disentangling) of possible correlations
between $CP$-odd and $CP$-even EFT operator effects in the light of the proposed
variables, a statistical combination of all 13~TeV data information possibly
merged to experimental low-energy data, as well as the impact on cosmology and
more precisely electroweak baryogenesis.

\section*{Acknowledgements}

The authors are grateful to K.~Mimasu and J.~No for discussions on $CP$
violation in models featuring an extended scalar sector, and to M.~Mangano and
A.~Nisati for guidance on the capabilities of the high-luminosity LHC project
on what concerns diboson production. We would like thank A.~Gritsan for pointing out many relevant references which were missing from the initial draft of the paper. FFF thanks the CNPq and the {\it Science
Without Borders} (grant SWE - 205350/2014-3) programs for supporting his visit
to Sussex University. This work has been supported in part by the Science
Technology and Facilities Council (STFC) under grant number ST/J000477/1, the
French ANR project DMAstroLHC (ANR-12-BS-005-0006) and the CNRS {\it Th\'eorie
LHC France initiative}. The work of DS  is supported by the National Science Foundation
under Grant PHY-1519045.

\appendix

\section{Event simulation and selection details}
\label{appA}

In order to simulate all LHC collision events required for this work, we have
used as a theoretical context the Standard Model Effective Field Theory
expressed in the
strongly interacting light Higgs basis~\cite{Giudice:2007fh,Contino:2013kra},
also known as the SILH basis. We have made used of the corresponding
implementation~\cite{Alloul:2013naa} in the {\sc FeynRules} package~\cite{%
Alloul:2013bka} to generate a UFO model~\cite{Degrande:2011ua} that we have
used within the {\sc MadGraph5\_aMC@NLO} platform~\cite{Alwall:2014hca}. We have
generated, for different choices of the EFT parameters, 150.000 hard scattering
events that we have then passed to {\sc Py\-thi\-a}~6~\cite{Sjostrand:2006za} for
parton showering and hadronization. The final state objects have been
reconstructed by employing the anti-$k_T$ algorithm~\cite{Cacciari:2008gp} with
an $R$-parameter set to 0.4 by using the {\sc FastJet}~\cite{Cacciari:2011ma}
interface of {\sc MadAnalysis}~5~\cite{Conte:2012fm,Conte:2014zja}. The latter
program has also been used to achieve all the analyses performed in this work,
after considering as $b$-tagged jets all jets for which a $B$-hadron is present
within a cone of radius $R=0.4$ centred on the jet momentum direction.

\subsection{$ZH$ associated production in the dilepton channel}
Reconstructed events are selected by demanding the presence of two isolated
leptons whose pseudorapidity satisfies $|\eta|<2.5$ and transverse momentum
$p_T$ is larger than 20~GeV. Moreover, we impose that the invariant mass of the
dilepton system is compatible with a $Z$-boson $m_{\ell\ell} \in [83, 110]$~GeV.
Lepton isolation is implemented by forbidding the presence of any
reconstructed object in a cone of radius $R = 0.4$ centred on the lepton
direction. We additionally request that the selected events
feature two $b$-tagged jets with a pseudorapidity $|\eta|<2.5$ and a transverse
momentum larger than 40~GeV and 20~GeV for the leading and subleading $b$-jet
respectively.

\subsection{$WH$ associated production in the single lepton channel}
We select events whose particle content features a single isolated charged
lepton
with a transverse momentum $p_T>10$~GeV and a pseudorapidity $|\eta|< 2.47$, two
$b$-tagged jets with a transverse momentum greater than 40~GeV and 20~GeV for
the leading and subleading jet respectively and with a pseudorapidity
$|\eta|<2.5$. Lepton isolation is implemented by forbidding the presence of any
reconstructed object in a cone of radius $R = 0.4$ centred on the lepton
direction.

\subsection{VBF Higgs boson production}
Events are selected by requiring the presence of two jets with a transverse
momentum \mbox{$p_{T}^{j}>20$~GeV}, a pseudrapidity $|\eta_{j}|<4.5$ and
typical VBF properties. The dijet invariant mass hence required to be larger
than 400~GeV and the jet separation in pseudrapidity is imposed to be above 2.8.

\subsection{$W$-boson pair production}
We select events featuring a final state with two isolated leptons whose
pseudorapidity satisfies $|\eta|<2.5$ and transverse momentum is larger than
20~GeV. Lepton isolation is implemented by forbidding the presence of any
reconstructed object lying in a cone of radius $R = 0.4$ centred on the lepton
direction, the jets candidate being jets with a transverse momentum larger than
20~GeV and a pseudorapidity smaller than 4.5 in absolute value.

\subsection{Higgs boson production and decay into the four-lepton channel}
Event selection relies on the presence of four isolated leptons with a
pseudorapidity $|\eta|<2.5$ and a transverse momentum $p_T>10$~GeV in the final
state. Jets candidate are defined
with a transverse momentum enforced to be larger than 20~GeV and a
pseudorapidity smaller than 4.5 in absolute value, and
lepton isolation is imposed by forbidding the presence of objects in a
cone of radius
$R = 0.4$ centred on the lepton direction.

\bibliographystyle{JHEP}
\bibliography{paper}

\end{document}